\documentclass[12pt]{article}

\usepackage{graphics}
\usepackage{color}
\usepackage{listings}
\usepackage{fullpage}
\usepackage[dvips]{lscape}
\usepackage[dvips]{graphics}
\usepackage{graphicx}
\usepackage{epsf}
\usepackage{epsfig}
\usepackage{rotating}
\usepackage{subfigure}
\usepackage{amsmath}
\usepackage{amssymb}
\usepackage{amsfonts}
\usepackage{mathrsfs}
\usepackage{latexsym}
\usepackage[export]{adjustbox}
\usepackage{authblk}
\usepackage{textpos}

\numberwithin{equation}{section}
\numberwithin{equation}{subsection}
\DeclareMathAlphabet{\mathpzc}{OT1}{pzc}{m}{it}
\DeclareMathOperator{\Tr}{Tr}

\begin{document}

\title{Entanglement Entropy and Variational Methods: \\[14pt] Interacting Scalar Fields \\[30pt]}

\author[1]{Jordan S. Cotler}
\affil[1]{Stanford Institute for Theoretical Physics, Department of Physics, Stanford University}

\author[2]{Mark T. Mueller}
\affil[2]{Center for Theoretical Physics, Department of Physics, Massachusetts Institute of Technology}

\renewcommand\Authands{ and }
\maketitle

\begin{textblock*}{5cm}(11cm,-12.2cm)
\fbox{\footnotesize MIT-CTP-4713}
\end{textblock*}

\begin{abstract}
\noindent

We develop a variational approximation to the entanglement entropy for scalar $\phi^4$ theory in 1+1, 2+1, and 3+1 dimensions, and then examine the entanglement entropy as a function of the coupling.  We find that in 1+1 and 2+1 dimensions, the entanglement entropy of $\phi^4$ theory as a function of coupling is monotonically decreasing and convex.  While $\phi^4$ theory with positive bare coupling in 3+1 dimensions is thought to lead to a trivial free theory, we analyze a version of $\phi^4$ with infinitesimal negative bare coupling, an asymptotically free theory known as \textit{precarious} $\phi^4$ theory, and explore the monotonicity and convexity of its entanglement entropy as a function of coupling.  Within the variational approximation, the stability of precarious $\phi^4$ theory is related to the sign of the first and second derivatives of the entanglement entropy with respect to the coupling.
\end{abstract}

\newpage
\tableofcontents

\newpage
\section{Introduction}
\label{sec:introduction}

Entanglement is a pervasive phenomenon in quantum physics, and has been studied extensively in finite-dimensional systems using the tools of quantum information theory.  More elusive is the entanglement structure of infinite-dimensional systems, namely quantum field theories.  The last twenty years has seen the development of a large literature on the entanglement entropy of QFT's and has shed light on the entanglement between spatial degrees of freedom \cite{CardyCalabrese1, CasiniHuerta1, CardyCalabrese2, TatsumaRyuTakayanagi1}.  Currently, there are only a few classes of field theories for which entanglement entropy can be computed exactly.  These include certain exactly solvable QFT's, free field theories, and conformal field theories.  CFT's with holographic duals are particularly interesting because their entanglement entropies can be computed using the Ryu-Takayanagi formula \cite{RyuTakayanagi1}. 

Despite this vast body of work, the entanglement entropy of QFT's remains rather mysterious.  Entanglement entropies for QFT's typically contain a combination of divergent and finite terms which depend on the parameters of the QFT.  Often it can be difficult to gain insight into the entanglement structure of a QFT by analyzing the form of its entanglement entropy.  More insightful is how the entanglement entropy \textit{changes} as a QFT undergoes dynamics such as a quench, or as parameters of the QFT are tuned.

Conspicuously missing from the list of QFT's with exactly computable entanglement entropies are those interacting field theories which comprise the Standard Model.  Even conventional QFT's such as $\phi^4$ theory and Yukawa theory have not had much presence in the entanglement entropy literature.  While it may not be possible to study the exact entanglement entropy of these theories, it is possible to study perturbative or variational approximations.

In this paper we embark on a program to analyze the entanglement entropies of more conventional interacting QFT's, and begin with interacting scalar field theories and $\phi^4$ theory in particular.  Our plan of attack is to use a variational principle to determine a non-perturbative approximation to the ground state of $\phi^4$ theory for arbitrary coupling, and then compute the entanglement entropy of the approximation.  Our variational ansatz will be a Gaussian wave functional, for which we can compute the entanglement entropy exactly.  Since the ground state of a massive free scalar field theory is exactly a Gaussian wave functional, our approximation will be accurate in a neighborhood of zero coupling.  We will also analyze the variational approximation at larger values of the coupling, for which the variational method is well-defined.  The validity and accuracy of the Gaussian variational approximation compares favorably to one-loop computations and large N approximations for theories where such comparisons make sense.  Our consideration of the approximation at larger values of the coupling will augment our understanding of the small coupling case, and provide evidence for features of entanglement entropy that may hold for all values of the coupling.

Once we have computed the variational approximation to the entanglement entropy in $\phi^4$ theory, we will analyze how it changes as we tune the coupling.  We will find that the derivatives of entanglement entropy with respect to coupling are often finite and independent of regularization, and represent meaningful physical quantities.  By studying the dependence of entanglement entropy on the coupling, we find several surprising features which suggest new insights into the nature of entanglement in QFT's.

The structure of the paper is as follows: First we discuss the variational methods used to approximate the ground state and other more general states of $\phi^4$ theory.  Then we explain how to compute the entanglement entropy of the variational ansatz using the replica trick.  Finally, we analyze the dependence of the entanglement entropy on the coupling in various dimensions for $\phi^4$ theory within the variational approximation.

\section{Variational Methods and the Gaussian Effective Potential}
\label{sec:vmgep}

In this section we introduce and review the essential aspects of the variational methods we will use for the approximate calculations of entanglement entropy in interacting quantum field theories.  The framework of variational methods that employ a Gaussian trial wave functional in the computation of the effective action and effective potential (in particular the 2PI effective action; see Cornwall, Jackiw, and Tomboulis \cite{CornwallJackiwTomboulis}) is referred to as the \emph{Gaussian Effective Potential}, or GEP.  The GEP for $\phi^4$ theory, as well as for other quantum field theories, has been studied extensively by Stevenson and his collaborators \cite{Stevenson0,Stevenson1,Stevenson2,Stevenson3} beginning in the 1980s.  That work built on earlier work of Barnes and Ghandour \cite{BarnesGhandour} and others in the 1970s, which in turn extended and clarified the pioneering work of Schiff \cite{Schiff} in the early 1960s. We will trace the development of the GEP and the closely related gap equation along a slightly different route, using a special case of a time-dependent variational principle based on work of Kerman and Koonin \cite{KermanKoonin}, and applied to quantum field theory by Jackiw, Kerman, and others \cite{JackiwKerman, Jackiw, PiSamiullah, KermanLin1, KermanLin2}.  In considering the special case of a static uniform system, our results reproduce those of Stevenson et al. for scalar $\phi^4$ theories.

%It is also worth mentioning how these variational methods are connected to the familiar language of Feynman diagrams.  It is well known that computing the one-loop one-particle irreducible (1PI) effective action is equivalent to a Hartree approximation, summing all single-loop diagrams (one loop, arbitrary number of external lines) [Coleman and Weinberg, Jackiw, standard QFT textbooks].  Perhaps less well-known but still well-established is that computing the two-particle irreducible (2PI) effective action, in a variational scheme connecting the one-point Green function (mean field) and the two-point Green function is equivalent to summing the infinite collection of so-called daisy and super-daisy graphs, consisting of all diagrams with loops upon loops upon loops \ldots. [Cornwall Jackiw Tomboulis, Barnes and Ghandour].  Indeed, it is possible to generalize further to a hierarchy of self-consistently determined n-point Green functions in the computation of the n-particle irreducible effective action [Brown - Whittingham], although we will not go to those lengths in this paper.

\subsection{The Variational Wave Functional and the Variational Hamiltonian}
\label{ssec:variational_wave_functional_variational_hamiltonian}

We will use the Schr\"odinger picture description of quantum field theory (see Jackiw \cite{Jackiw} for an excellent review and additional references), where quantum field operators $\hat{\phi}(\mathbf{x})$ and their canonical momenta $\hat{\pi}(\mathbf{x})$ are time-independent, and the wave functional $\Psi[\phi(\mathbf{x}),t] = \langle \phi(\mathbf{x}) | \Psi(t) \rangle$ is in general time-dependent, although we will not need to consider the full time-dependence in this paper.  Our goal is to approximately describe the quantum field theory using variational methods.  We define the variational Hamiltonian functional $\bar{H}$ by
\begin{equation}
	\bar{H} = \langle \Psi | \hat{H} | \Psi \rangle
\end{equation}
where $\hat{H}$ is the Hamiltonian operator for the quantum field theory.  The variational method consists of minimizing this variational Hamiltonian functional with respect to a set of variables which parameterize the wave functional $\Psi$.  We make the ansatz that the variational wave functional is a Gaussian, parameterized by a real-valued function $\bar{\phi}(\mathbf{x})$ and a real-valued symmetric kernel $G(\mathbf{x},\mathbf{x'})$
\begin{equation*}
	\Psi[\phi] = N \exp \Big[ - \frac{1}{4} \int d^d\mathbf{x} \int d^d\mathbf{x'}
	\left( \phi(\mathbf{x}) - \bar{\phi}(\mathbf{x}) \right) G^{-1}(\mathbf{x},\mathbf{x'}) 
	\left( \phi(\mathbf{x'}) - \bar{\phi}(\mathbf{x'}) \right) \Big]
\end{equation*}
where $d$ is the spatial dimension.  The kernel $G(\mathbf{x},\mathbf{x'})$ must be positive-definite in order for the variational wave function to be normalizable, and choosing $N = [\det(2\pi G)]^{-1/4}$ assures that $\Psi$ has unit norm.  We will frequently describe the kernel $G(\mathbf{x},\mathbf{x'})$ as matrix elements of an operator $G$ acting on the space of single-particle wave functions.  For example, in position space, $G(\mathbf{x},\mathbf{x'}) = ( \mathbf{x} | \, G \, | \, \mathbf{x'} )$.  It is important to distinguish the state space of the quantum field theory where the states $| \Psi \rangle$ are defined and upon which the quantum field operators $\hat{\phi}(\mathbf{x})$ and $\hat{\pi}(\mathbf{x})$ act, and the space of single-particle wave functions where the functions $\bar{\phi}$ and $\bar{\pi}$ are defined and upon which operators such as $G$ and $G^{-1}$ act.  Using the more compact operator notation, we may write the variational wave functional as
\begin{equation}
	\Psi[\phi] = [\det(2\pi G)]^{-1/4} \exp \Big[ \! - \frac{1}{4} \, ( \phi - \bar{\phi}) \, G^{-1} ( \phi - \bar{\phi}) \Big]
\end{equation}
As it stands, this variational wave functional is a formal expression that will be given more concrete meaning through the regularization and renormalization procedures discussed in subsequent sections.

Although the vacuum state of the theory corresponds to $\bar{\phi} = 0$, other values of the scalar field lead to a variety of physically meaningful and interesting results.  Considering states other than the vacuum is necessary to compute (the variational approximation to) the effective action for the quantum field theory.  Similar wave functionals are considered in \cite{BenedictPi}, where those authors were motivated by studying entanglement entropy in the context of free field theory in 1+1 dimensions.

The Schr\"odinger picture field operators $\hat{\phi}(\mathbf{x})$ and their canonically conjugate momentum operators $\hat{\pi}(\mathbf{x})$ act on the wave functional $\Psi[\phi,t]$ in the canonical way
\begin{align}
	\hat{\phi}(\mathbf{x}) \Psi[\phi] &= \phi(\mathbf{x)} \Psi[\phi]  \\
	\hat{\pi}(\mathbf{x}) \Psi[\phi] &= - i \frac{\delta}{\delta \phi(\mathbf{x})} \Psi[\phi]
\end{align}
It follows immediately that the variational parameters $\bar{\phi}(\mathbf{x})$ are the expectation values of the Schr\"odinger picture operator in the Gaussian variational state $| \Psi \rangle$
\begin{equation}
	\bar{\phi}(\mathbf{x},t) = \langle \Psi(t) | \, \hat{\phi}(\mathbf{x}) \, | \Psi(t) \rangle
\end{equation}
and we can also immediately see that $G(\mathbf{x},\mathbf{x'})$ is the two-point function or Green function
\begin{equation}
	G(\mathbf{x},\mathbf{x'}) 
	= \langle \Psi(t) | \, [\hat{\phi}(\mathbf{x}) - \bar{\phi}(\mathbf{x})] \, [\hat{\phi}(\mathbf{x'}) - \bar{\phi}(\mathbf{x'})] \, | \Psi \rangle
\end{equation}

We now focus on a specific interacting scalar quantum field theory using the variational method just discussed.  The Hamiltonian operator for this quantum field theory is given by
\begin{equation}
	\hat{H} = \int d^d\mathbf{x} \left[ \, \frac{1}{2} \, \hat{\pi}^2 + \frac{1}{2} \, (\nabla \hat{\phi})^2 + V(\hat{\phi})  \, \right]
\end{equation}
where $V$ is the scalar field potential given in terms of the bare parameters of the theory.  We will be concerned with a quartic potential of the form
\begin{equation}
	V(\phi) = \frac{m_B^2}{2} \, \phi^2 + \frac{\lambda_B}{4!} \, \phi^4
\end{equation}
where $m_B$ is a bare mass parameter and $\lambda_B$ is a bare coupling parameter.  The coupling $\lambda_B$ has mass dimension 2, 1, and 0 in 1+1, 2+1, and 3+1 dimensions, respectively.  Using the Gaussian variational wave functional $\Psi$ and the Hamiltonian operator $\hat{H}$, a straightforward calculation gives
\begin{equation}
	\big\langle \Psi \big| \int d^d\mathbf{x} \frac{1}{2} \hat{\pi}^2(\mathbf{x}) \big| \Psi \big\rangle 
	= \frac{1}{8} \text{Tr} \,G^{-1}
	= \frac{1}{8} \int d^d\mathbf{x} \, (\mathbf{x} | G^{-1} |\mathbf{x}) 
	= \frac{1}{8} \int d^d\mathbf{x} \, G^{-1}(\mathbf{x},\mathbf{x})
\end{equation}
For the gradient-squared term in the Hamiltonian, we find
\begin{equation}
	\big\langle \Psi \big| \int d^d\mathbf{x} \, \frac{1}{2} ( \nabla \hat{\phi}(\mathbf{x}) )^2 \, \big| \Psi \big\rangle 
		= \int d^d\mathbf{x} \, \frac{1}{2} \left( \nabla \bar{\phi}(\mathbf{x}) \right)^2 
		+ \frac{1}{2} \text{Tr} \, \mathbf{p}^2 G
\end{equation}
where $\mathbf{p}$ is the momentum operator on the space of single-particle wavefunctions, given by $(\mathbf{x} | \, \mathbf{p} \, | \mathbf{x'}) = - i \nabla_{\mathbf{x}} \delta^d(\mathbf{x} - \mathbf{x'})$ in the position basis.
Finally, the expectation value of the potential operator $V(\hat{\phi})$ is defined by
\begin{equation}
	\overline{V}(\bar{\phi},\overline{G}) = \langle \Psi(t) | \, V(\hat{\phi}) \, | \Psi(t) \rangle
\end{equation}
where, for the quartic potential, a short computation (equivalent to applying Wick's theorem) shows that
\begin{equation}
	\overline{V}(\bar{\phi},\overline{G}) = \frac{m_B^2}{2} \, ( \bar{\phi}^{\, 2} + \overline{G} \, )
	+ \frac{\lambda_B}{4!}\, ( \bar{\phi}^{\, 4} + 6 \, \bar{\phi}^{\, 2} \, \overline{G} + 3 \, \overline{G}^{\, 2} \, )
\end{equation}
where we have defined $\overline{G}(\mathbf{x}) = (\mathbf{x} | \, G \, | \mathbf{x}) = G(\mathbf{x},\mathbf{x})$.  Only this position-space-diagonal part of the Green function appears in $\overline{V}$, because in a local quantum field theory all of the Schr\"odinger picture field operators $\hat{\phi}(\mathbf{x})$ appearing in the Hamiltonian operator $\hat{H}$ are evaluated at the same point in space.  As in any continuum quantum field theory we expect $\overline{G}(\mathbf{x})$ to be divergent, which we will address in subsequent discussions of renormalization.  It is worth emphasizing that the last term in the above expression is the square of a matrix element $\overline{G}^{\, 2} \equiv (\mathbf{x} | \, G \, | \mathbf{x})^2 = G(\mathbf{x},\mathbf{x})^2$, \emph{not} the matrix element of the square of the operator $G$, which we would write as $(\mathbf{x} | \, G^2 \, | \mathbf{x}) = G^2(\mathbf{x},\mathbf{x})$.  Combining the results above, the full variational Hamiltonian is
\begin{equation}
	\bar{H} = \text{Tr} \, \left[ \frac{1}{8} \, G^{-1} + \frac{1}{2} \, \mathbf{p}^2 G \, \right]
	+ \int d^d\mathbf{x} \, \left[ \frac{1}{2} (\nabla \bar{\phi})^2 + \overline{V}(\bar{\phi},\overline{G}) \, \right]
\end{equation}
The equations for the variational parameters, for an arbitrary potential function $V$, follow directly from minimizing this functional with respect to $\bar{\phi}$ and $G$
\begin{align}
	\frac{\delta \bar{H}}{\delta \bar{\phi}} &= 0 \quad\quad \Longrightarrow \quad\quad
	- \nabla^2 \bar{\phi} + \frac{\partial \overline{V}}{\partial \bar{\phi}} = 0  \\[8pt]
	\frac{\delta \bar{H}}{\delta G} &= 0 \quad\quad \Longrightarrow \quad\quad\quad
	\frac{1}{4} \, G^{-2} = \omega^2
\end{align}
where the last equation is written in the condensed operator form.  The operator $\omega$ is defined by its kernel in position space as
\begin{equation}
	\omega^2(\mathbf{x},\mathbf{x'}) = \left( - \nabla^2_{\mathbf{x}} + m^2 \right) \delta^d(\mathbf{x} - \mathbf{x'})
\end{equation}
where the quantity $m(\bar{\phi},\overline{G})$ is defined, for an arbitrary potential function $V$, by 
\begin{equation}
	m^2 = 2 \, \frac{\partial \overline{V}}{\partial \overline{G}}
\end{equation}
and for the quartic potential in particular, it is
\begin{equation}
	m^2 = 2 \, \frac{\partial \overline{V}}{\partial \overline{G}} = m_B^2 + \frac{\lambda_B}{2} \, ( \bar{\phi}^{\, 2} + \overline{G} \, )
\end{equation}
The quantity $m$ will play a significant role in subsequent developments.  In addition to depending on the bare mass $m_B$ and bare coupling $\lambda_B$, $m$ is a function of $\bar{\phi}(\mathbf{x})$ and $\overline{G}(\mathbf{x})$, so it may depend on position in general.  In the case of a uniform system, $m$ will be a constant parameter.  We now turn our attention to such static uniform systems.

\subsection{The Gap Equation and the Gaussian Effective Potential}
\label{ssec:gap_equation_and_GEP}

If we now restrict ourselves to spatially uniform systems, which are translation invariant, then $\bar{\phi}$ does not depend on position and $\nabla^2 \bar{\phi} = 0$.  Additionally, the Green function $G(\mathbf{x},\mathbf{x'})$ will only depend on the difference $\mathbf{x} - \mathbf{x'}$.  Because of translation invariance, the analysis simplifies considerably in momentum space.  Defining the Fourier transforms
\begin{align}
	G(\mathbf{x},\mathbf{x'}) 
	&= \int \frac{d^d\mathbf{p}}{(2\pi)^d} \, \tilde{G}(\mathbf{p}) \, e^{i \mathbf{p} \cdot (\mathbf{x} - \mathbf{x'})}  \\[4pt]
	\omega(\mathbf{x},\mathbf{x'}) 
	&= \int \frac{d^d\mathbf{p}}{(2\pi)^d} \, \tilde{\omega}(\mathbf{p}) \, e^{i \mathbf{p} \cdot (\mathbf{x} - \mathbf{x'})}
\end{align}
and using the definition of of $\omega$, its Fourier transform takes the simple form
\begin{equation}
	\tilde{\omega}(\mathbf{p},m) = \sqrt{\mathbf{p}^2 + m^2}
\end{equation}
where we now indicate explicitly the dependence $m$.  The second of the two non-trivial Hamilton equations, for $G^{-2}$ in terms of $\omega^2$, is now
\begin{equation}
	\frac{1}{4} \, \tilde{G}(\mathbf{p},m)^{-2} = \tilde{\omega}(\mathbf{p},m)^2 = \mathbf{p}^2 + m^2
\end{equation}
This important equation, together with
\begin{equation}
\label{gapequation0}
	m^2 = m_B^2 + \frac{\lambda_B}{2} \, [ \, \bar{\phi}^{\, 2} + \overline{G}(m) \, ]
\end{equation}
is called the \emph{gap equation}.  The expression for $\overline{G}$, now given in terms of $m$, is
\begin{equation}
	\overline{G}(m) = G(\mathbf{x},\mathbf{x}) = \int \frac{d^d\mathbf{p}}{(2\pi)^d} \, \tilde{G}(\mathbf{p},m)
	= \int \frac{d^d\mathbf{p}}{(2\pi)^d \, 2 \tilde{\omega}(\mathbf{p},m)}
\end{equation}
The gap equation gives the functional form of the Green function $G$ in the static uniform case \emph{and} an equation connecting the mean field $\bar{\phi}$ to the mass parameter $m$.  We will frequently refer to just this relationship between $\bar{\phi}$ and $m$ in (\ref{gapequation0}) as the gap equation, although the full meaning is as just discussed.  The Green function describes a particle in the interacting quantum field theory propagating with a mass $m$.  We have arrived at this interpretation of the Green function $G$ and the mass $m$ via the variational principle and the Gaussian form of the variational wave functional; it is not a separate assumption.

The variational Hamiltonian simplifies in the uniform case to
\begin{equation}
	\bar{H} = \text{Tr} \, \left[ \, \frac{1}{8} \, G^{-1} + \frac{1}{2} \, \mathbf{p}^2 G \, \right]
	+ \int d^d\mathbf{x} \, \overline{V}(\bar{\phi},\overline{G})
\end{equation}
Computing the trace in the momentum space representation, the variational Hamiltonian density $\bar{\mathcal{H}}$, defined by $\bar{H} = \int d^d\mathbf{x} \, \bar{\mathcal{H}}$, is given by
\begin{equation}
	\bar{\mathcal{H}}(\bar{\phi},G)
	= \int \frac{d^d\mathbf{p}}{(2\pi)^d} 
	\left[ \frac{1}{8} \, \tilde{G}(\mathbf{p})^{-1} + \frac{1}{2} \, \mathbf{p}^2 \tilde{G}(\mathbf{p}) \, \right]
	+ \overline{V}(\bar{\phi},\overline{G})
\end{equation}
If we look back through the derivation of this expression, we see that the first term, with $G^{-1}$, comes from the expectation value of the $\hat{\pi}^2$ term of the Hamiltonian operator; the second term, with $\mathbf{p}^2 G$, comes from the expectation value of the $(\nabla\hat{\phi})^2$ term of the Hamiltonian operator; and the third term $\overline{V}$ comes from the expectation value of the potential $V(\hat{\phi})$ term of the Hamiltonian operator.  The sum of the first two terms is less divergent than either one separately, which is a consequence of the Lorentz invariance of the theory.  We now use the relationship of the momentum space Green function $\tilde{G}(\mathbf{p},m)$ to $\tilde{\omega}(\mathbf{p},m)$ via the gap equation, so the momentum space integrals may be written as
\begin{align*}
	\int \frac{d^d\mathbf{p}}{(2\pi)^d} 
	&\left[ \, \frac{1}{8} \, \tilde{G}(\mathbf{p},m)^{-1} + \frac{1}{2} \, \mathbf{p}^2 \tilde{G}(\mathbf{p},m) \, \right]
	= \int \frac{d^d\mathbf{p}}{(2\pi)^d} 
		\left[ \, \frac{1}{4} \, \tilde{G}(\mathbf{p},m)^{-1} - \frac{m^2 }{2} \, \tilde{G}(\mathbf{p},m) \, \right]  \\[6pt]
	= &\int \frac{d^d\mathbf{p}}{(2\pi)^d \, 2 \tilde{\omega}(\mathbf{p},m)} 
		\left[ \, \tilde{\omega}(\mathbf{p},m)^2 - \frac{m^2}{2} \right]
	= I_1(m) - \frac{m^2}{2} \, I_0(m)
\end{align*}
where, following \cite{Stevenson2}, we have defined the integrals
\begin{equation}
	I_N(m) = \int \frac{d^d\mathbf{p}}{(2\pi)^d \, 2 \tilde{\omega}(\mathbf{p},m)} \ \tilde{\omega}(\mathbf{p},m)^{2N}
	= \frac{1}{2} \int \frac{d^d\mathbf{p}}{(2\pi)^d} \ (\mathbf{p}^2 + m^2)^{N - \frac{1}{2}}
\end{equation}
We note in particular that $\overline{G}(m) = I_0(m)$ is the integral appearing in the gap equation and the $\overline{V}$ part of the variational Hamiltonian.  Most of these integrals will be divergent in the cases we will be studying, and the renormalization prescriptions we are about to discuss will address this in detail.  

The variational Hamiltonian density $\bar{\mathcal{H}}$, with $m$ given as a function of the mean field $\bar{\phi}$ via the gap equation, is known as the \emph{Gaussian Effective Potential} (GEP), and is given by
\begin{equation}
	V_{GE}(\bar{\phi}) = I_1(m) - \frac{m^2}{2} I_0(m) + \overline{V}
\end{equation}
and $\overline{V}$ for the quartic potential was given earlier by
\begin{equation}
	\overline{V} = \frac{m_B^2}{2} ( \bar{\phi}^{\, 2} + I_0(m) )
	+ \frac{\lambda_B}{4!}\, ( \bar{\phi}^{\, 4} + 6 \, \bar{\phi}^{\, 2} I_0(m) + 3 \, I_0(m)^2 )
\end{equation}
where we have used $\overline{G}(m) = I_0(m)$.

We now turn to the renormalization of the theory and a physical interpretation of the formal expressions for the gap equation and the GEP.

\subsection{Renormalization of the Gaussian Effective Potential}

The vacuum state, with $\bar{\phi} = 0$, corresponds to a minimum of $V_{GE}(\bar{\phi})$
\begin{equation}
	\frac{d V_{GE}}{d \bar{\phi}} \bigg\vert_{\bar{\phi} = 0} = 0
\end{equation}
Following Stevenson \cite{Stevenson2} as well as Barnes and Ghandour \cite{BarnesGhandour}, we define the renormalized mass $m_R$
\begin{equation}
	m^2_R = \frac{d^2 V_{GE}}{d \bar{\phi}^2} \bigg\vert_{\bar{\phi} = 0}
\end{equation}
and the renormalized coupling $\lambda_R$
\begin{equation}
	\lambda_R = \frac{d^4 V_{GE}}{d \bar{\phi}^4} \bigg\vert_{\bar{\phi} = 0}
\end{equation}
Computing these derivatives is routine, and the results are quite simple when the derivatives are evaluated in the vacuum state $\bar{\phi} = 0$.  For the renormalized mass $m_R$ we have
\begin{equation}
	m^2_R = m_B^2 + \frac{\lambda_B}{2} \, I_0(m_0) = m_0^2
\end{equation}
The second equality above follows because $m_0$ satisfies the gap equation $m_0^2 = m_B^2 + \frac{\lambda_B}{2} \, I_0(m_0)$ with $\bar{\phi} = 0$ .  So, the renormalized mass $m_R$ is identified with the vacuum solution of the gap equation: $m_R = m_0$.  The renormalized mass sets the scale for all dimensionful quantities in the renormalized theory, and it will often be convenient to work in units such that $m_R = 1$.  For the renormalized coupling $\lambda_R$, we have
\begin{equation}
	\lambda_R = \lambda_B \, \frac{1 - \lambda_B I_{-1}(m_R) / 2}{1 + \lambda_B I_{-1}(m_R) / 4}
\end{equation}
With these renormalization prescriptions, we can now derive the renormalized forms of the gap equation and the GEP in various dimensions.

The detailed discussion of the 1+1 and 2+1 dimensional cases are quite similar.  In these dimensions there are ultraviolet divergences that require mass renormalization, but coupling renormalization is merely a finite coupling reparameterization, and in these lower dimensions we will express results in terms of the original finite bare coupling $\lambda_B$.  We refer the reader to the papers of Stevenson and collaborators for more detail, but we will review and summarize the most important results here.  In 1+1 and 2+1 dimensions, we will make use of the following identities for the divergent integrals $I_1(m)$ and $I_0(m)$\,:
\begin{align}
	I_1(m) - I_1(m_R) &= \frac{1}{2} (m^2 - m_R^2) \, I_0(m_R) + F_1(m)  \\[4pt]
	I_0(m) - I_0(m_R) &= F_0(m)
\end{align}
where the finite quantities $F_0(m)$ and $F_1(m)$ are
\begin{align}
	F_1(m) &= - \frac{1}{8\pi} ( m^2 \log \frac{m^2}{m_R^2} - m^2 + 1 )	\quad\quad\quad \ \	 \text{1 + 1 dimensions}  \\[4pt]
	F_1(m) &= -\frac{1}{24\pi} \, (m - m_R)^2 (2 m + m_R)			\quad\quad\quad 	 \text{2 + 1 dimensions}
\end{align}	
and
\begin{align}
	F_0(m) &= - \frac{1}{4\pi} \log \frac{m^2}{m_R^2}	\quad\quad\quad\quad\quad 		\text{1 + 1 dimensions}  \\[4pt]
	F_0(m) &= - \frac{1}{4\pi} (m - m_R)				\quad\quad\quad\quad 		\text{2 + 1 dimensions}
\end{align}
The renormalization of the gap equation proceeds in essentially the same way in 1+1 and 2+1 dimensions.  Starting with the bare gap equation and using the definition of the renormalized mass, we have
\begin{align}
	m^2 	&= m_B^2 + \frac{\lambda_B}{2} I_0(m) + \frac{\lambda_B}{2} \bar{\phi}^2  \\[4pt]
	&= m_B^2 + \frac{\lambda_B}{2} I_0(m_R) + \frac{\lambda_B}{2} (I_0(m) - I_0(m_R)) 
		+ \frac{\lambda_B}{2} \bar{\phi}^2  \\[4pt]
	&= m_R^2 + \frac{\lambda_B}{2} F_0(m) + \frac{\lambda_B}{2} \bar{\phi}^2
\end{align}
As a check, note that when $\bar{\phi}$ takes its vacuum value $\bar{\phi} = 0$, the mass parameter $m$ also takes its vacuum value $m = m_0 = m_R$, since $F_0(m_R) = 0$.  So, the renormalized gap equations for each dimension are
\begin{align}
	\bar{\phi}^2 &= \frac{1}{4\pi} \log \frac{m^2}{m_R^2} + \frac{2}{\lambda_B} (m^2 - m_R^2)
	\quad\quad\quad\quad\quad\quad \,	\text{1 + 1 dimensions}  \\[4pt]
	\bar{\phi}^2 &= \frac{1}{4\pi} (m - m_R) + \frac{2}{\lambda_B} (m^2 - m_R^2)
	\quad\quad\quad\quad\quad 	\text{2 + 1 dimensions}
\end{align}
In either case, the renormalized gap equation is of the form
\begin{equation}
	\bar{\phi}^2 =  - F_0(m) + \frac{2}{\lambda_B} (m^2 - m_R^2)
\end{equation}
We see the similarity of the form of the gap equations across different dimensions.  Only the first term has a different form in different dimensions and has no dependence on the form of the potential in the original Hamiltonian.  The last term contains the coupling $\lambda_B$ and depends on the detailed form of the potential.  As we will see, a very similar result holds in 3+1 dimensions, where the gap equation will be expressed in terms of the renormalized coupling instead of the bare coupling.

The renormalization of the Gaussian Effective Potential $V_{GE}$ also proceeds in essentially the same way in 1+1 and 2+1 dimensions.  Consider the first two terms of $V_{GE}$\,:
\begin{align*}
	I_1(m) - &\frac{m^2}{2} I_0(m) = I_1(m_R) + I_1(m) - I_1(m_R) - \frac{m^2}{2} I_0(m)  \\[6pt]
	&= I_1(m_R) + \frac{1}{2} (m^2 - m_R^2) I_0(m_R) + F_1(m) - \frac{m^2}{2} I_0(m)  \\[6pt]
	&= I_1(m_R) - \frac{m_R^2}{2} \, I_0(m_R)  + F_1(m) - \frac{m^2}{2} (I_0(m) - I_0(m_R))  \\[6pt]
	&= I_1(m_R) - \frac{m_R^2}{2} \, I_0(m_R)  + F_1(m) - \frac{m^2}{2} F_0(m)
\end{align*}
The first two terms contribute an additive divergent constant to $V_{GE}$, but we will require the energy density to be zero in the vacuum state, using the freedom to add a constant to the original energy density to make this so.  Using the above expressions for $F_0$ and $F_1$, the finite parts are quite simple:
\begin{align}
	F_1(m) - \frac{m^2}{2} F_0(m) &= \frac{1}{8\pi} (m^2 - m_R^2 ) 		\quad\quad\quad \ \ \ \ \text{1 + 1 dimensions}  \\[6pt]
	F_1(m) - \frac{m^2}{2} F_0(m) &= \frac{1}{24\pi} (m^3 - m_R^3)		\quad\quad\quad \ \	\text{2 + 1 dimensions}
\end{align}	
Turning to the third and final term of $V_{GE}$, namely $\overline{V}$, a useful form is given by (recall that $\overline{G}(m) = I_0(m)$)
\begin{align*}
	&\overline{V}(\bar{\phi},\overline{G})
	= \frac{m_B^2}{2} \, ( \bar{\phi}^{\, 2} + \overline{G}(m) ) 
	+ \frac{\lambda_B}{4!} \, ( \bar{\phi}^{\, 4} + 6 \, \bar{\phi}^{\, 2} \, \overline{G}(m) + 3 \, \overline{G}(m)^2 )  \\[4pt]
	&= \frac{\lambda_B}{4!} \, \bar{\phi}^4 + \frac{1}{2} ( m_B^2 + \frac{\lambda_B}{2} \, \overline{G}(m) ) \bar{\phi}^2
	+ \frac{1}{2 \lambda_B} ( m_B^2 + \frac{\lambda_B}{2} \, \overline{G}(m) )^2 - \frac{m_B^4}{2 \lambda_B}  \\[4pt]
	&= \frac{\lambda_B}{4!} \, \bar{\phi}^4 + \frac{1}{2} ( m^2 - \frac{\lambda_B}{2} \bar{\phi}^2 ) \, \bar{\phi}^2
		+ \frac{1}{2 \lambda_B} ( m^2 - \frac{\lambda_B}{2} \bar{\phi}^2 )^2 - \frac{m_B^4}{2 \lambda_B}  \\[4pt]
		\label{Vbar1}
	&= -\frac{\lambda_B}{12} \, \bar{\phi}^4 + \frac{m^4}{2 \lambda_B} - \frac{m_B^4}{2 \lambda_B}
\end{align*}
where we have once again used the bare gap equation.  This expression (\ref{Vbar1}) holds in any dimension.  Note that the last term above is also an additive divergent constant.  Combining the results above for the two parts of $V_{GE}$, and choosing the overall additive constant such that $V_{GE} = 0$ in the vacuum state characterized by $\bar{\phi} = 0$ and $m = m_0 = m_R$, we have the final result for the renormalized GEP in each dimension
\begin{align}
	V_{GE}(\bar{\phi}) &= \frac{1}{8\pi}(m^2 - m_R^2) + \frac{1}{2 \lambda_B} (m^4 - m_R^4) - \frac{\lambda_B}{12} \bar{\phi}^4
	\quad\quad \ \,		\text{1 + 1 dimensions}  \\[4pt]
	V_{GE}(\bar{\phi}) &=  \frac{1}{24\pi}(m^3 - m_R^3) + \frac{1}{2 \lambda_B} (m^4 - m_R^4) - \frac{\lambda_B}{12} \bar{\phi}^4
	\quad\quad 		\text{2 + 1 dimensions}
\end{align}
As already noted for the gap equations, this way of writing things clearly shows the similarity of the GEP across different dimensions.  Only the first term has a different form in different dimensions; it arises purely from expectation value of the operators $\hat{\pi}^2$ and $(\nabla\hat{\phi})^2$ terms in the Hamiltonian operator $\hat{H}$, and has no dependence on the form of the potential operator $V(\hat{\phi})$.  The last two terms, which contain the coupling $\lambda_B$, depend only on the choice of potential.  Of course, the dependence of $m$ on $\bar{\phi}$ is given by the appropriate gap equation in each dimension, and these have similar but not identical forms.

In 3+1 dimensions, we will restrict our attention to a version of $\phi^4$ theory which is non-trivial in 3+1 dimensions -- meaning it is physically well-defined and is not a free field theory.  This is the theory which Stevenson \cite{Stevenson2} has given the name \emph{precarious} $\phi^4$ theory.  In this theory, the bare coupling is taken to be \emph{negative} and arbitrarily close to zero in a limit in which the ultraviolet cutoff goes to infinity.  This theory, while unstable, is in a very specific sense arbitrarily metastable as the ultraviolet cutoff goes to infinity, and constitutes a physically sensible effective field theory.  We do not attempt to review precarious $\phi^4$ theory and refer the reader to the very clear discussion in Stevenson \cite{Stevenson2}.  As noted there, and observed much earlier by Symanzik \cite{Symanzik} and Parisi \cite{Parisi}, this version of $\phi^4$ theory has the intriguing property of being asymptotically free.  We do not consider the ``standard'' version of $\phi^4$ theory with positive bare coupling because it is trivial using the variational methods of the GEP -- meaning it is a free field theory with the renormalized coupling driven to zero from above as the ultraviolet cutoff goes to infinity.  This is in harmony with other ``triviality" results for standard $\phi^4$ theory in 3+1 (and higher) spacetime dimensions.  
According to Stevenson and Tarrach \cite{StevensonTarrach, StevensonDC}, in addition to the precarious version there is one other physically sensible version of $\phi^4$ theory in 3+1 dimensions which has a positive bare coupling but differs from the standard theory.  They call this theory \emph{autonomous} $\phi^4$ theory.  We will not consider this version in this paper.

In 3+1 dimensions the ultraviolet divergences require coupling renormalization as well as mass renormalization.  The prescription for the renormalized coupling in any dimension, as formulated earlier, is
\begin{equation}
	\label{eqn:coupling_renormalization}
	\lambda_R = \frac{d^4 V_{GE}}{d \bar{\phi}^4} \bigg\vert_{\bar{\phi}_0 = 0}
	= \lambda_B \, \frac{1 - \lambda_B I_{-1}(m_R) / 2}{1 + \lambda_B I_{-1}(m_R) / 4}
\end{equation}
In the precarious $\phi^4$ theory, since the bare coupling $\lambda_B$ is negative and driven to zero as the cutoff goes to infinity,  the coupling renormalization is given by
\begin{equation}
	\frac{\lambda_B I\, _{-1}(m_R)}{4} = -1 - 3 \, \frac{4}{\lambda_R \, I_{-1}(m_R)} + \cdots
\end{equation}
where we have dropped terms of order $I_{-1}(m_R)^{-2}$ or smaller.  The following identities for the integrals $I_N(m)$ in 3+1 dimensions are useful for the renormalization procedure:
\begin{align}
	I_1(m) - &I_1(m_R) = \frac{1}{2} \, (m^2 - m_R^2) \, I_0(m_R) - \frac{1}{8} (m^2 - m_R^2)^2 \, I_{-1}(m_R) + F_1(m)  \\
	I_0&(m) - I_0(m_R) = - \frac{1}{2} (m^2 - m_R^2) \, I_{-1}(m_R) + F_0(m)  \\[4pt]
	&\quad I_{-1}(m) - I_{-1}(m_R) = F_{-1}(m)
\end{align}
where the finite parts are
\begin{align}
	F_1(m) &= \frac{1}{128\pi^2} \, ( m^4 \log \frac{m^4}{m_R^4} - 3(m^2 - m_R^2)^2 - 2 m_R^2 (m^2 - m_R^2) )  \\[4pt]
	F_0(m) &= - \frac{1}{16\pi^2} (m^2 - m_R^2 - m^2 \log \frac{m^2}{m_R^2})  \\[4pt]
	F_{-1}(m) &= - \frac{1}{8\pi^2} \log \frac{m^2}{m_R^2}
\end{align}
Using these identities and the coupling renormalization, it is now straightforward to derive the renormalized gap equation, which is
\begin{equation}
	\bar{\phi}^2 = \frac{1}{16\pi^2} \, ( m^2 - m_R^2 - m^2 \log \frac{m^2}{m_R^2} ) + \frac{6}{\lambda_R} (m^2 - m_R^2) 
\end{equation}
Following similar derivations already discussed for lower dimensions, a straightforward calculation shows that the renormalized GEP is
\begin{equation}
	V_{GE}(\bar{\phi}) = \frac{1}{128\pi^2} \, (m^4 - m_R^4 - m^4 \log \frac{m^4}{m_R^4}) + \frac{3}{2\lambda_R} (m^4 - m_R^4)
\end{equation}
It is worth noting that there is a non-trivial cancellation of logarithmic divergences between the kinetic and potential pieces of the Hamiltonian, associated with the coupling renormalization.  This is the key to the renormalizability of the theory in 3+1 dimensions.  It does not occur and is not required in lower dimensions.  

We now express all dimensionful quantities in units of the renormalized mass $m_R$, so that $m_R = 1$.  It is convenient to define the normalized coupling $\alpha = \lambda_R/96\pi^2$ and the negative normalized inverse coupling $\kappa = -1/\alpha$, and a re-scaled version of the scalar field by $\varphi = 4\pi \bar{\phi}$ so that $\varphi^2 = 16\pi^2 \bar{\phi}^2$.  Then the renormalized gap equation takes the form
\begin{equation}
	\varphi^2 = ( (\kappa - 1)(1 - m^2) - m^2 \log m^2 )  \\[4pt]
\end{equation}
and the renormalized GEP is
\begin{equation}
	V_{GE} = \frac{1}{128\pi^2} \, ( (2 \kappa - 1)(1 - m^4) - m^4 \log m^4 )
\end{equation}
Following \cite{Stevenson2}, for each value of the coupling we define the \emph{critical value} $\varphi_c$ as the largest value of the scalar field where the gap equation has a solution.  These are given by
\begin{align}
	\varphi_c^2 &= \kappa - 1 + \exp (-\kappa)  \\
   	 m_c^2 &= \exp (-\kappa)
\end{align}
The \emph{break value} $\varphi_b$ is defined as the largest value of $\bar{\phi}$ where the solution of the gap equation gives the global minimum of the GEP.  When $\varphi > \varphi_b$, the GEP is a constant $V_{GE}(\varphi) = V_{GE}(\varphi_b)$, corresponding to the minimizing value of $m = 0$.
\begin{align}
	\varphi_b^2 &= \kappa - 1 + \frac{1}{2} \exp \left(\frac{1}{2} -\kappa\right)  \\
	m_b^2 &= \exp \left(\frac{1}{2} -\kappa \right)
\end{align}
It is always the case that $\varphi_b < \varphi_c$ and $m_b > m_c$.  The region between these values corresponds to a metastable region of the GEP, where the solution to the gap equation corresponds to a local but not global minimum.  As discussed in detail in \cite{Stevenson2}, when the coupling is strong enough, specifically when $\kappa \le 1/2$, the break value $\varphi_b$ goes to zero and the GEP becomes flat for all values of $\varphi$, corresponding to an interacting theory with a massless particle.  Keeping track of the critical values and break values will be important in the discussion of our results for the entanglement entropy in 3+1 dimensions.

We need to make some final observations before we move on to entanglement entropy, concerning the sign of the coupling in each dimension.  Once again, Stevenson \cite{Stevenson2} provides a clear discussion of these issues, and we merely summarize his findings.  In 1+1 and 2+1 dimensions, the concern is over infrared divergences.  In these cases it is essential that the bare coupling be non-negative: $\lambda_B \ge 0 $.  Otherwise, the Gaussian Effective Potential is either completely ill-defined (in 1+1) or is unbounded below (in 2+1).  In 3+1 dimensions for precarious $\phi^4$ theory the concern is not infrared divergences and a GEP unbounded below, but ultraviolet divergences that require coupling renormalization.  However, something even more interesting takes place: there is a duality relation that connects positive renormalized couplings to negative renormalized couplings, along with a rescaling of the mass parameter $m$ and the scalar field $\bar{\phi}$, and with an additive constant shift of the GEP.  The explicit formulas defining the duality transformation are
\begin{equation}
\label{duality1}
	m^2_* = m^2 / F  \quad\quad\quad\quad
	\bar{\phi}^2_* = \bar{\phi}^2 / F  \quad\quad\quad\quad
	\kappa_* = \kappa + \log F
\end{equation}
where the asterisk denotes the dual values, and the scale factor $F$ is the solution of
\begin{equation}
\label{duality2}
	(F - 1)(1 - \kappa) = F \log F
\end{equation}
The entire range of positive couplings $\alpha > 0$ (or equivalently negative inverse couplings $\kappa < 0$) gets mapped to a range of negative couplings $\alpha < -1$ (or equivalently a range of positive inverse couplings $0 < \kappa < 1$) under duality, so the duality mapping is one-to-one but not onto.  Note the somewhat counter-intuitive results that $\kappa = 0$ is the self-dual point, indicating that the two formulations with infinitely negative coupling and infinitely positive coupling are dual to each other, and that arbitrarily weak positive coupling is \emph{not} dual to arbitrarily weak negative coupling.  This duality relation means we need only consider negative values of the renormalized coupling $\alpha$ with no loss of generality.  In later sections on entanglement entropy we will express our results in terms of the negative renormalized couplings, although we will briefly discuss how some of these results look for positive couplings which are dual to the negative couplings for purposes of comparison with results in lower dimensions.

On the following pages we plot the solutions of the renormalized gap equation and the renormalized Gaussian Effective Potential in 1+1, 2+1, and 3+1 dimensions, for different values of the coupling.  We also plot the duality relations for coupling and scale.  We emphasize that all dimensionful quantities in these plots are measured in units of the renormalized mass scale $m_R$, which sets the scale for the entire theory.  In these units, $m_R = 1$.

\begin{figure}[hp]
	\begin{centering}
	\scalebox{1.0}{\includegraphics[width=1.0\textwidth,center]{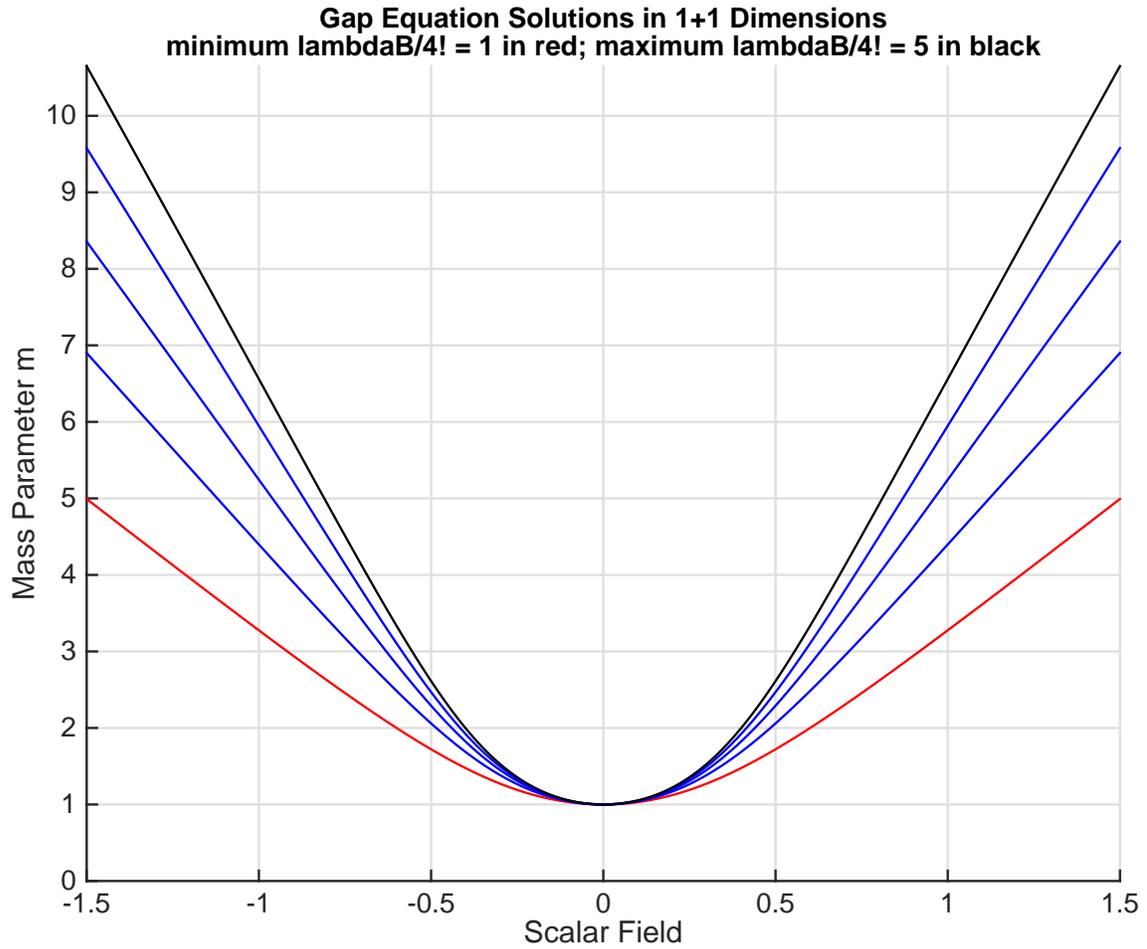}}
	\end{centering}
	\caption{Solutions of the renormalized gap equation in 1+1 Dimensions, with a range of values for the bare coupling $\lambda_B$.  The bottom curve (red in color plot) corresponds to the weakest coupling; the top curve (black in color plot) corresponds to the strongest coupling.}
	\label{fig:gap_equation_solutions_1d}
\end{figure}

\begin{figure}[hp]
	\begin{centering}
	\scalebox{1.0}{\includegraphics[width=1.0\textwidth,center]{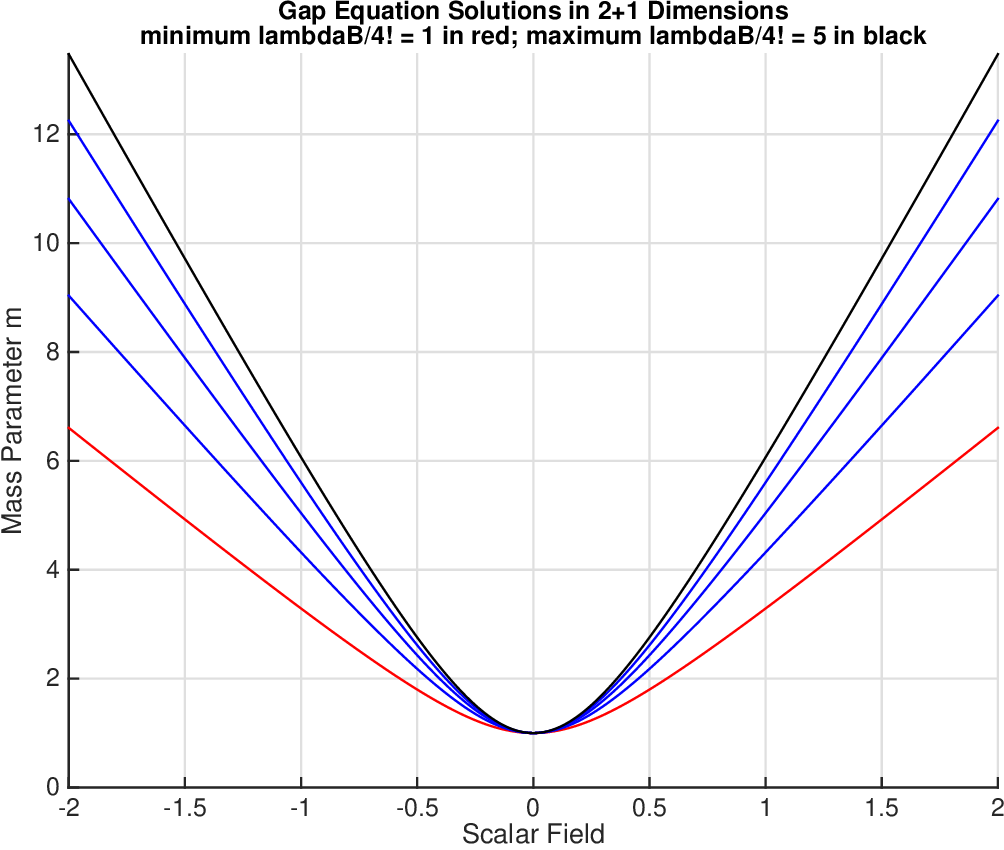}}
	\end{centering}
	\caption{Solutions of the renormalized gap equation in 2+1 Dimensions, with a range of values for the bare coupling $\lambda_B$.    The bottom curve (red in color plot) corresponds to the weakest coupling; the top curve (black in color plot) corresponds to the strongest coupling.}
	\label{fig:gap_equation_solutions_2d}
\end{figure}

\begin{figure}[hp]
	\begin{centering}
	\scalebox{1.0}{\includegraphics[width=1.0\textwidth,center]
	{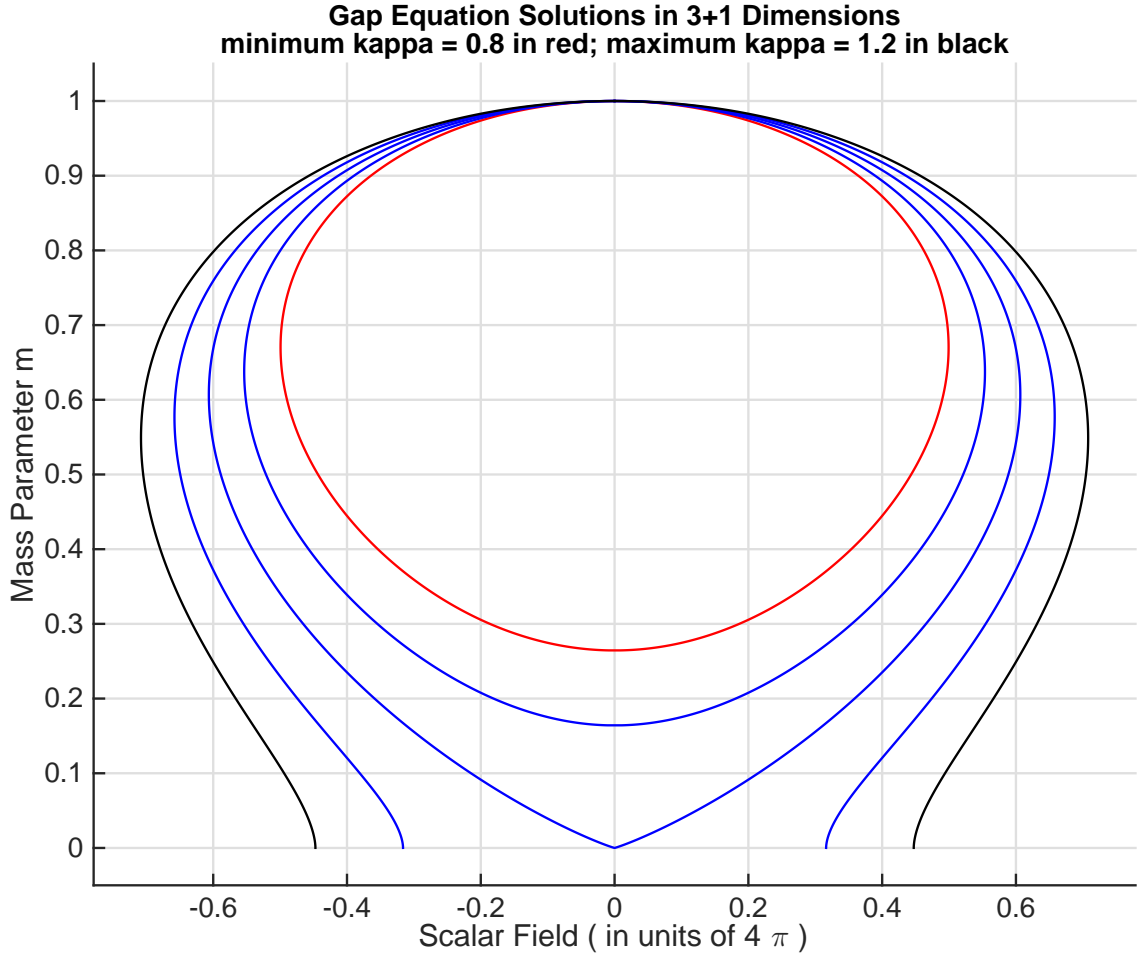}}
	\end{centering}
	\caption{Solutions of the renormalized gap equation in 3+1 Dimensions, with a range of values for the inverse coupling $\kappa$.  The innermost curve (red in color plot) corresponds to the strongest coupling (most negative $\alpha$ or equivalently smallest $\kappa$); The outermost curve (black in color plot) corresponds to the weakest coupling (least negative $\alpha$ or equivalently largest $\kappa$).}
	\label{fig:gap_equation_solutions_3d}
\end{figure}

\begin{figure}[hp]
	\begin{centering}
	\scalebox{1.0}{\includegraphics[width=1.0\textwidth,center]{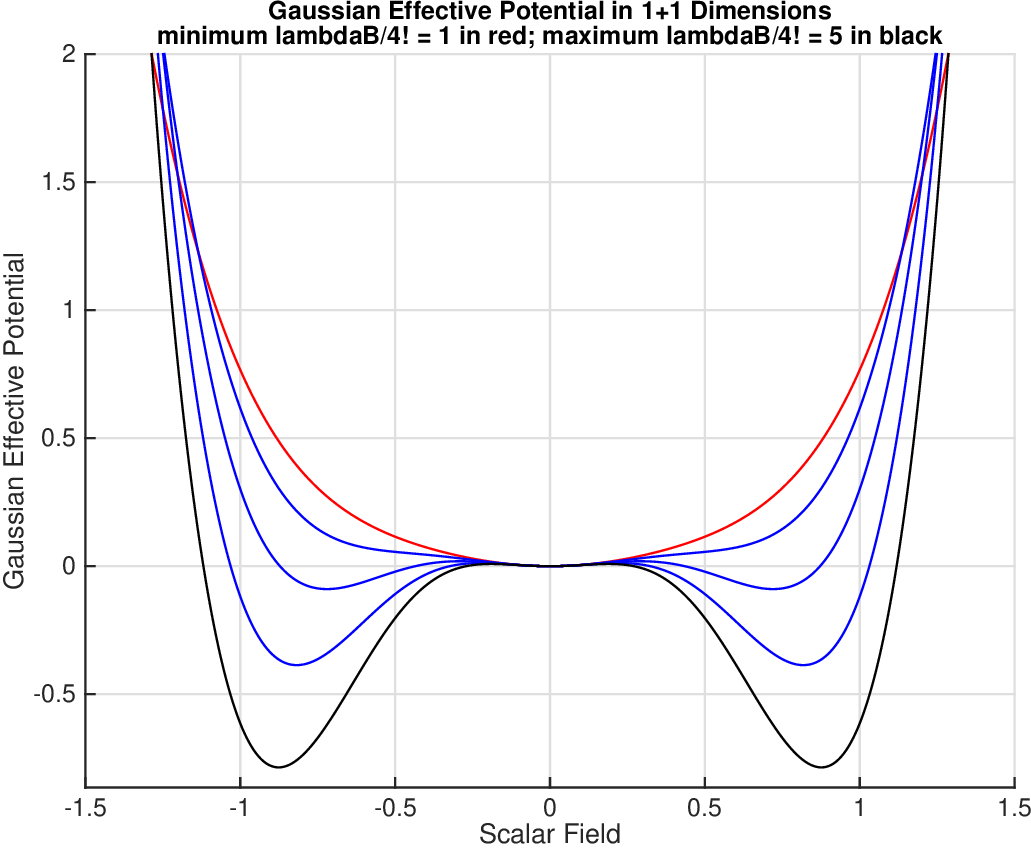}}
	\end{centering}
	\caption{Renormalized Gaussian Effective Potential in 1+1 Dimensions, with a range of values for the bare coupling $\lambda_B$.  The top curve (red in color plot) corresponds to the weakest coupling; the bottom curve (black in color plot) corresponds to the strongest coupling.}
	\label{fig:gaussian_effective_potential_1d}
\end{figure}

\begin{figure}[hp]
	\begin{centering}
	\scalebox{1.0}{\includegraphics[width=1.0\textwidth,center]{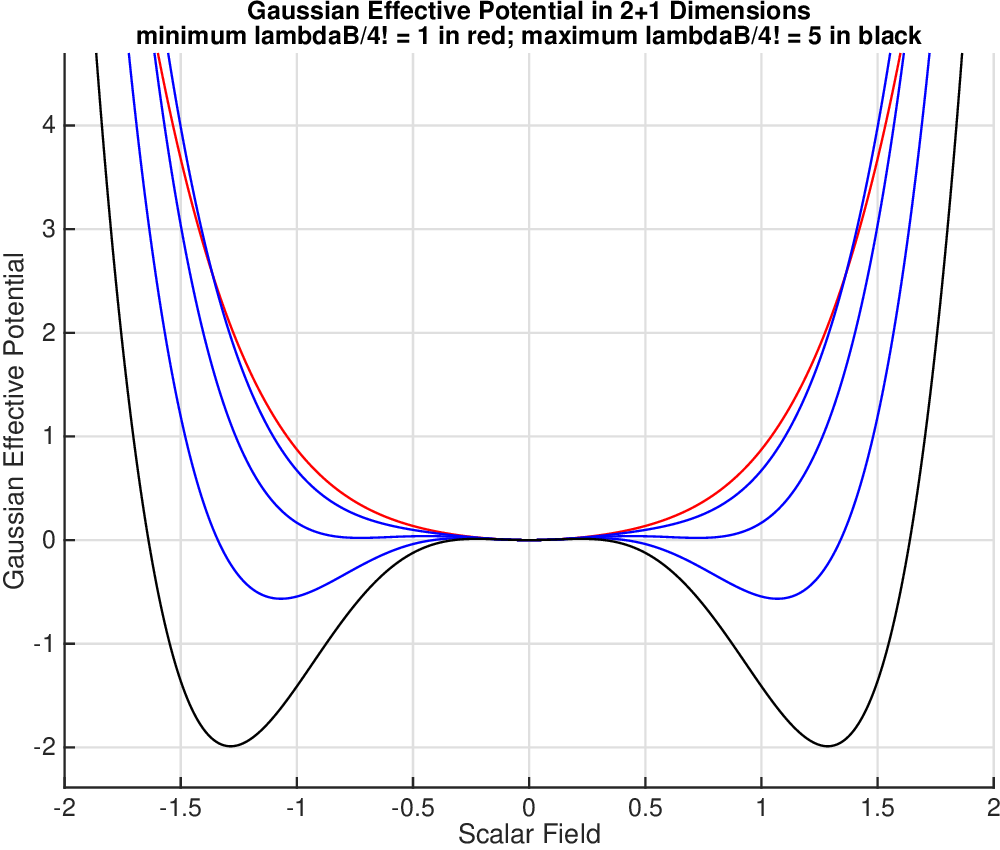}}
	\end{centering}
	\caption{Renormalized Gaussian Effective Potential in 2+1 Dimensions, with a range of values for the bare coupling $\lambda_B$.  The top curve (red in color plot) corresponds to the weakest coupling; the bottom curve (black in color plot) corresponds to the strongest coupling.}
	\label{fig:gaussian_effective_potential_2d}
\end{figure}

\begin{figure}[hp]
	\begin{centering}
	\scalebox{1.0}{\includegraphics[width=1.0\textwidth,center]{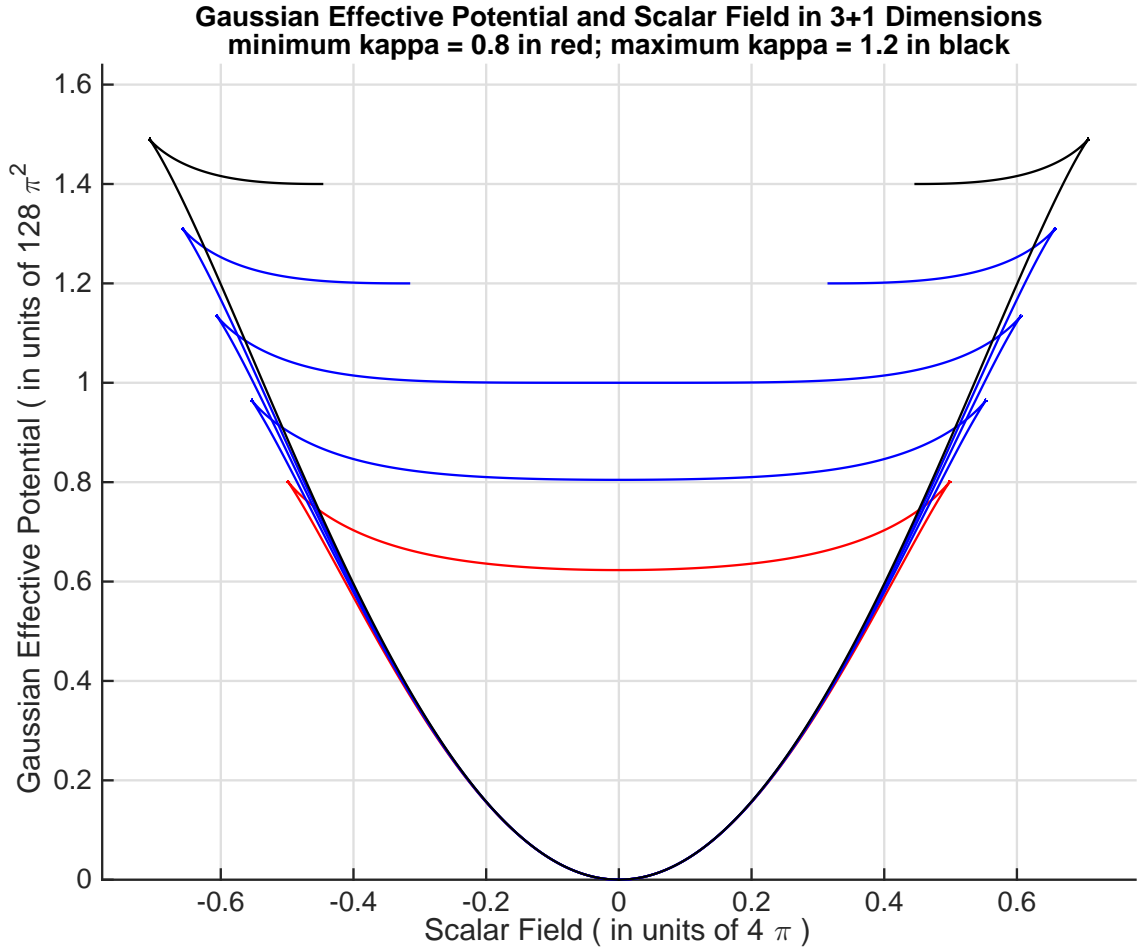}}
	\end{centering}
	\caption{Renormalized Gaussian Effective Potential in 3+1 Dimensions, with a range of inverse negative couplings $\kappa$. The bottom curve (red in color plot) corresponds to the strongest coupling (most negative $\alpha$ or equivalently smallest $\kappa$); The top curve (black in color plot) corresponds to the weakest coupling (least negative $\alpha$ or equivalently largest $\kappa$).}
	\label{fig:gaussian_effective_potential_3d}
\end{figure}

\begin{figure}[hp]
	\begin{centering}
	\scalebox{1.0}{\includegraphics[width=1.0\textwidth,center]{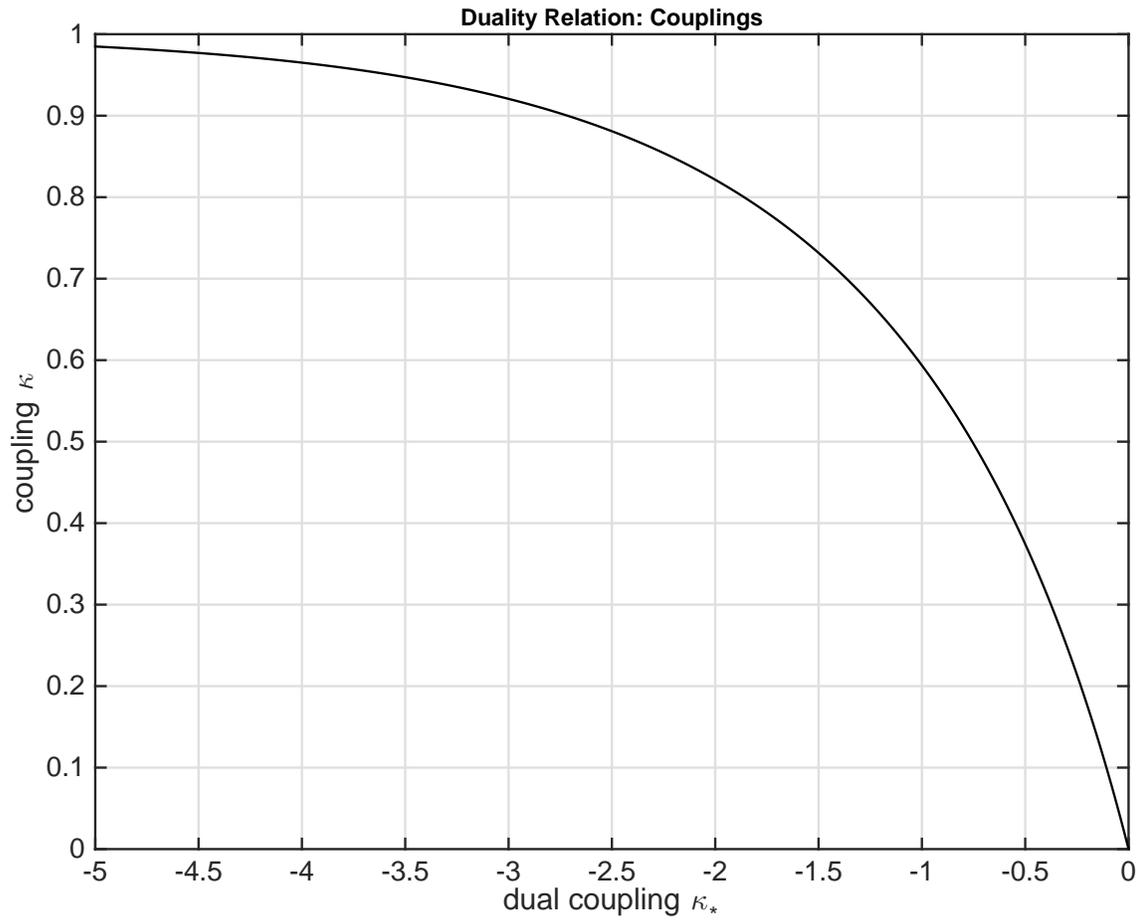}}
	\end{centering}
	\caption{Duality relation for precarious $\phi^4$: negative inverse coupling $\kappa \ge 0$ as a function of dual inverse coupling $\kappa_* \le 0$.  The duality transformations are given in equations \eqref{duality1} and \eqref{duality2}.}
	\label{fig:duality_coupling}
\end{figure}

\begin{figure}[hp]
	\begin{centering}
	\scalebox{1.0}{\includegraphics[width=1.0\textwidth,center]{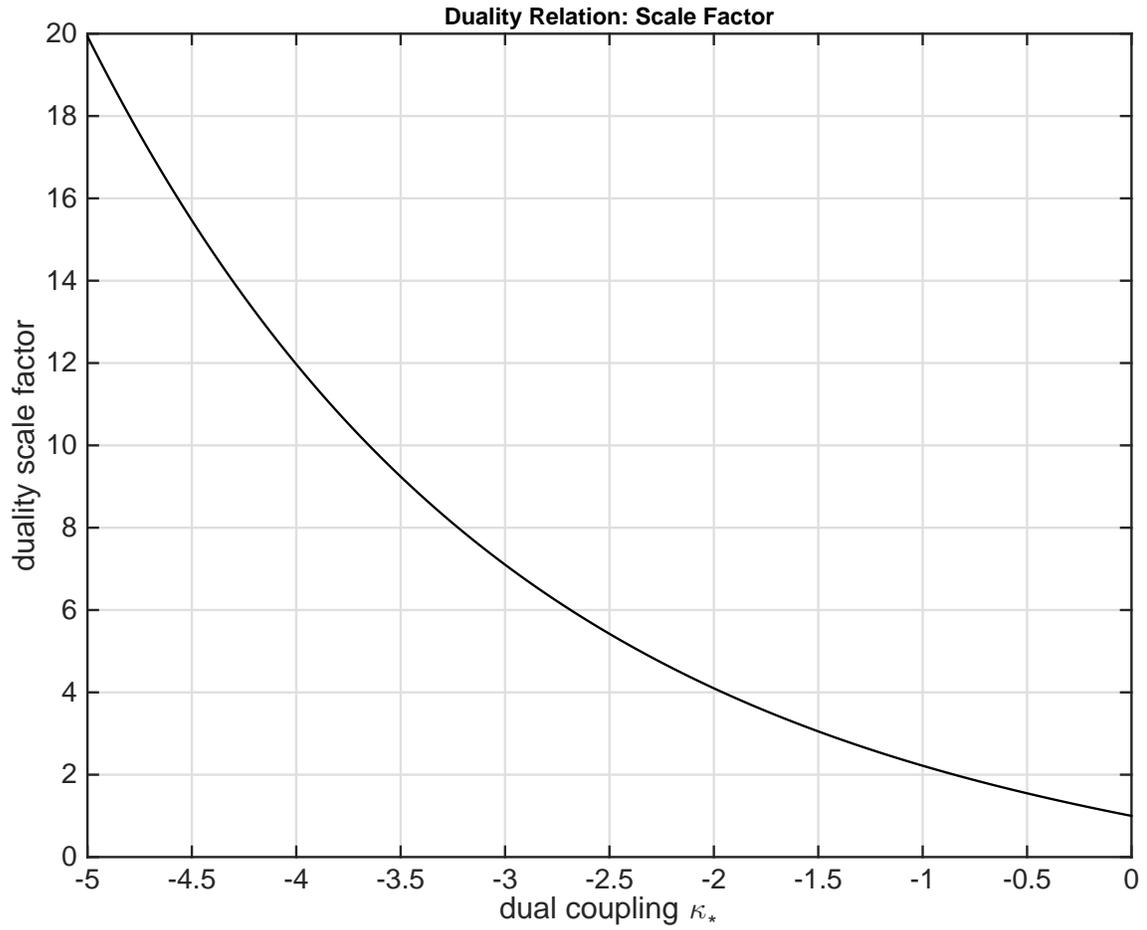}}
	\end{centering}
	\caption{Duality relation for precarious $\phi^4$: duality scale factor $F$ as a function of dual inverse coupling $\kappa_* \le 0$.  The duality transformations are given in equations \eqref{duality1} and \eqref{duality2}.}
	\label{fig:duality_scale}
\end{figure}

\newpage
\section{Entanglement Entropy from the Replica Trick}
\label{sec:hkee}

Since our variational ansatz for $\phi^4$ theory takes the same form as the ground state of a massive free scalar field, we would like to know how to compute the entanglement entropy of a massive free scalar field.  We will use the replica trick, which provides an efficient way of computing the von Neumann entropy of a density matrix.  Let $\rho$ denote an un-normalized density matrix, and let $\hat{\rho}$ denote the corresponding normalized density matrix $\hat{\rho} := \rho/\Tr(\rho)$.  Then the replica trick \cite{CallanWilczek1} gives us
\begin{equation}
\label{replica1}
\lim_{n \to 1}\left( - \frac{d}{dn} + 1 \right) \log(\Tr \rho^n) = - \Tr(\hat{\rho} \log \hat{\rho})
\end{equation}
equation (\ref{replica1}) is particularly useful since it normalizes our density matrix of interest, $\rho$.  This allows us not to worry about the normalization of our density matrix prior to applying the replica trick.

With the help of the replica trick, computing the entanglement entropy is conceptually straightforward but technically difficult.  Given a wave functional $\Psi[\phi(\textbf{x})]$ of our quantum field theory, we start by tracing out the degrees of freedom in a spacetime region $\mathcal{V}$ from the density matrix $\Psi[\phi(\textbf{x})] \Psi^*[\phi(\textbf{x})]$ to obtain the reduced density matrix $\rho$, which need not be normalized.  Then we compute $\Tr \rho^n$, and use the replica trick to obtain the desired entanglement entropy.  

In our case, we consider the Gaussian wave functional discussed in detail earlier
\begin{equation*}
	\Psi[\phi] = [\det(2\pi G)]^{-1/4} \exp \Big[ \! - \frac{1}{4} \, \phi \, G^{-1} \, \phi \Big]
\end{equation*}
Although we are actually interested in wave functionals which take the form \newline $\widetilde{\Psi}[\phi(\textbf{x})] = \Psi[\phi(\textbf{x}) + \bar{\phi}]$, letting $\widetilde{\rho}$ be the reduced density matrix of  $\widetilde{\Psi}[\phi(\textbf{x})] \widetilde{\Psi}^*[\phi(\textbf{x})] = \Psi[\phi(\textbf{x}) - \bar{\phi} ] \Psi^*[\phi(\textbf{x}) - \bar{\phi}]$, we see that $\Tr \rho^n = \Tr \widetilde{\rho}^n$ by a change of variables.  Therefore, it suffices to consider the entanglement entropy corresponding to $\Psi[\phi]$.  In the context of free field theories, the entanglement entropy for such Gaussian states has been considered by a number of authors, originating in the work of Sorkin and collaborators \cite{Sorkin, SorkinEtAlia}.  

For this paper, we will restrict ourselves to the case in which $\mathcal{V}$ is the half-space.  When $\mathcal{V}$ is the half-space, it can be shown that $\Tr \rho^n$ is proportional to the partition function $Z_\delta$ of a massive free scalar field living on an $n$-sheeted covering of the Euclidean plane, which is a cone of deficit angle $\delta = 2 \pi (1-n)$.  This deficit angle is non-positive because each power of the density matrix adds another sheet which contributes $2\pi$ radians of angle ``surplus."  Since $\frac{d}{dn} = - 2 \pi \frac{d}{d\delta}$, when $\mathcal{V}$ is the half-space equation (\ref{replica1}) can be written as \cite{CallanWilczek1, Solo1}
\begin{equation}
\label{deficit1}
\lim_{\delta \to 0} \left(2 \pi \frac{d}{d\delta} + 1\right) \log Z_{\delta} = - \Tr(\hat{\rho} \log \hat{\rho})
\end{equation}
Using heat kernel methods, it is possible to compute $\log Z_{\delta}$, which will give us the leading area law contributions to the entanglement entropy \cite{HertzbergWilczek1, Solo1, Dowker1}.  Using equation (\ref{deficit1}) we find that
\begin{equation}
\label{EE}
	S = \frac{A}{12 (4 \pi)^{(d-1)/2}} \int_{\epsilon^2}^\infty \frac{ds}{s^{(d+1)/2}} \, e^{- m^2 s} 
	\quad\quad\quad \ \	 \text{d + 1 dimensions}  \\[4pt]
\end{equation}
where $\epsilon$ is the length scale which serves as our UV cutoff, $\Sigma$ is the boundary of $\mathcal{V}$, and $A$ is the area of $\Sigma$.  The expansion of equation (\ref{EE}) contains divergent terms due to the UV cutoff.
%Since \textit{any} smooth surface $\Sigma'$ looks locally like the boundary of the half-space (i.e., a hyperplane), by summing over local ``pieces" of entanglement entropy, the $\frac{1}{\epsilon^{d-1}}$ term in the expansion of Eqn. (\ref{EE}) provides the leading contribution to the entanglement entropy for generic volumes $\mathcal{V}$ with smooth boundary $\Sigma$ \cite{Solo1}.
In this paper, we will be interested in the cases $d=1,2,3$.  For $d=1$, we have
\begin{align}
	S &= - \frac{A}{12} \,\text{Ei}\left(-m^2 \epsilon ^2\right)  \nonumber  \\[4pt]
 	&= - \frac{A}{12} \left(\log(m^2 \epsilon^2) + \gamma_E \right) + \mathcal{O}(m^2 \epsilon^2)  \nonumber  \\[4pt]
	&= - \frac{A}{12} \left( \log \varepsilon^2 + \log m^2 \right)
		+ \mathcal{O}(m^2 \epsilon^2)  \quad\quad\quad\quad \ \	 \text{1 + 1 dimensions}
\end{align}
where $\text{Ei}(z)$ is the exponential integral and $\gamma_E$ is the Euler-Mascheroni constant, and without loss of generality we have rescaled $\epsilon$ to $\varepsilon = \epsilon \, e^{\gamma_E/2}$.  In $d=1$ dimensions only, we have that $A=1$.  For $d=2$,
\begin{align}
	S &= \frac{A}{12} \left(\frac{e^{-m^2 \epsilon ^2}}{\sqrt{\pi} \epsilon } - m + m\,  \text{erf}(m \epsilon ) \right) \nonumber \\
   	&= \frac{A}{12} \left(\frac{1}{\sqrt{\pi} \epsilon} - m \right) + \mathcal{O}(m^2 \epsilon) 
	\quad\quad\quad\quad\quad\quad\quad\quad	 \text{2 + 1 dimensions} 
\end{align}
where $\text{erf}(z)$ is the error function.  Finally, for $d=3$,
\begin{align}
	S &= \frac{A}{48 \pi} \left(\frac{e^{-m^2 \epsilon ^2}}{\epsilon ^2} 
		+ m^2 \, \text{Ei}\left(-m^2 \epsilon ^2\right) \right) \nonumber \\
	&= \frac{A}{48 \pi} \left(\frac{1}{\epsilon^2} + m^2 \log \epsilon^2 + m^2 \log m^2 + m^2(\gamma_E - 1) \right) 
		+ \mathcal{O}(m \epsilon) \nonumber \\
	&= \frac{A}{48 \pi} \left(\frac{e^{\gamma_E}}{\varepsilon^2} + m^2 \log \varepsilon^2 + m^2 \log m^2 - m^2 \right) 
		+ \mathcal{O}(m \varepsilon) \quad\quad \ \	 \text{3 + 1 dimensions}
\end{align}
where, as in 1+1 dimensions, we have again rescaled $\epsilon$ to $\varepsilon = \epsilon \, e^{\gamma_E/2}$.

We note that there are other sub-leading area terms which can be obtained using more sophisticated methods \cite{Casinietal1}, but they will not affect our analysis.  In particular, when we consider derivatives of the entanglement entropy with respect to the coupling, these other sub-leading area terms will either vanish in the $\epsilon \to 0$ limit, or if they diverge will be sub-leading to other divergences. 

One may notice that the divergences in $S$ in $d$ spatial dimensions take the same form as the divergences of the effective potential for the massive free scalar field in $d$ spatial dimensions.  Indeed, this is no coincidence.  Noting that $W_{\delta} = -\log Z_{\delta}$ is the effective action for a free massive scalar field with deficit angle $\delta$, we can rewrite equation (\ref{deficit1}) as
\begin{equation}
\label{deficit2}
\lim_{\delta \to 0} \left(- 2 \pi \frac{d}{d\delta} - 1\right) W_{\delta} = - \Tr(\hat{\rho} \log \hat{\rho})
\end{equation}
Since the effective action contains the effective potential and thus the divergences of the effective potential, it is unsurprising that the entanglement entropy also contains divergences of the same form.  In fact, the entanglement entropy takes the following simple unifying form in all dimensions:
\begin{equation}
\label{unifyingform1}
	S / A = \frac{\pi}{3} \, \overline{G}(m) = \frac{\pi}{3} I_0(m)
\end{equation}
where we recall that $\overline{G}(m) = G(\mathbf{x},\mathbf{x})$ is the divergent diagonal piece of the Green function in position space.  Let us define $\Delta S$ by the finite part of the entanglement entropy $S$ plus an additive constant which makes $\Delta S$ equal to zero in the free field vacuum.  Then in all dimensions we have
\begin{equation}
\label{unifyingform2}
	\Delta S / A = \frac{\pi}{3} \, F_0(m)
\end{equation}
For our variational approximation, we will simply replace $m$ in the above equations with the variational mass.

%It may seem unnatural to consider only the finite part of the entanglement entropy $\Delta S$ since the divergent pieces contain information about the UV limit of the QFT.  However, the divergent pieces depend on the QFT's UV completion whereas the finite or ``universal" pieces do not.  From our the perspective of this paper, there is an alternate reason why $\Delta S$ is the physically relevant quantity to study.  We are primarily interested in how the entanglement entropy of a theory depends on the coupling.

%$S(\alpha) - S(0) = \Delta S + \text{const.}$

%where the constant term is finite.

\section{Calculations of Entanglement Entropy}
\label{sec:vcee}

\subsection{Entanglement Entropy Per Unit Area}
\label{ssec:EE_per_unit_area}

In this section, we will compute the entanglement entropy of $\phi^4$ theory within the variational approximation, and examine how the entanglement entropy depends on the coupling.  In a preceding section on the variational method and Gaussian Effective Potential, we found that the renormalized gap equations in spatial dimensions $d = 1, 2, 3$ take very similar forms, which we repeat here:
\begin{align}
	\bar{\phi}^2 &= \frac{1}{4\pi} \log \frac{m^2}{m_R^2} + \frac{2}{\lambda_B} (m^2 - m_R^2)
	\quad\quad\quad\quad\quad\quad\quad\quad\quad\quad\quad \,
	\text{1 + 1 dimensions}  \\[4pt]
	\bar{\phi}^2 &= \frac{1}{4\pi} (m - m_R) + \frac{2}{\lambda_B} (m^2 - m_R^2)
	\quad\quad\quad\quad\quad\quad\quad\quad\quad\quad
	\text{2 + 1 dimensions}  \\[4pt]
	\bar{\phi}^2 &= - \frac{1}{16\pi^2} \, (m_R^2 - m^2 + m^2 \log \frac{m^2}{m_R^2} ) - \frac{6}{\lambda_R} (m_R^2 - m^2)
	\quad\quad
	\text{3 + 1 dimensions}
\end{align}
Also, in the previous section on the formalism of entanglement entropy, we found that expressions for $\Delta S$ in spatial dimensions $d = 1, 2, 3$ also take very similar forms, expressed in all cases by $\Delta S/A = \frac{\pi}{3} F_0(m)$\,:
\begin{align}
	\Delta S / A &= -\frac{1}{12} \log \frac{m^2}{m_R^2}
	\quad\quad\quad\quad\quad\quad\quad\quad\quad\quad\quad \,
	\text{1 + 1 dimensions}  \\[4pt]
	\Delta S / A &= -\frac{1}{12} \, (m - m_R)
	\quad\quad\quad\quad\quad\quad\quad\quad\quad\quad
	\text{2 + 1 dimensions}  \\[4pt]
	\Delta S / A &= -\frac{1}{48\pi} \, ( m^2 - m_R^2 - m^2 \log \frac{m^2}{m_R^2} \, )
	\quad\quad\quad 	
	\text{3 + 1 dimensions}
\end{align}
To put all of these equations in the exactly the same form is now just a question of normalization, so we define the normalized dimensionless coupling $\alpha$ and the normalized dimensionless scalar field $\varphi$ for each dimension as
\begin{align}
	\alpha &= \frac{\lambda_B}{8\pi m_R^2}
	\quad\quad\quad\quad\quad
	\varphi^2 = 4 \pi \bar{\phi}^2
	\quad\quad\quad\quad\quad \
	\text{1 + 1 dimensions}  \\[4pt]
	\alpha &= \frac{\lambda_B}{8\pi m_R}
	\quad\quad\quad\quad\quad
	\varphi^2 = 4 \pi \bar{\phi}^2
	\quad\quad\quad\quad\quad \
	\text{2 + 1 dimensions}  \\[4pt]
	\alpha &= \frac{\lambda_R}{96\pi^2}
	\quad\quad\quad\quad\quad \ \,
	\varphi^2 = 16 \pi^2 \bar{\phi}^2
	\quad\quad\quad\quad \ \,
	\text{3 + 1 dimensions}
\end{align}
and the normalized dimensionless quantities appearing in the gap equation
\begin{align}
	f &= \log m^2 
	\quad\quad\quad\quad\quad\quad\quad\quad\quad\quad \
	\text{1 + 1 dimensions}  \\[4pt]
	f &= m - 1 
	\quad\quad\quad\quad\quad\quad\quad\quad\quad\quad \
	\text{2 + 1 dimensions}  \\[4pt]
	f &= - ( 1 - m^2 + m^2 \log m^2 ) 
	\quad\quad\quad
	\text{3 + 1 dimensions}
\end{align}
where $m$ and other dimensionful parameters are now expressed in units of the renormalized mass scale $m_R$, so in effect we have chosen units such that $m_R = 1$.  Finally, define the normalized entanglement entropy per unit area $\sigma$ by
\begin{align}
	\sigma = 12 \, \Delta S / A
	\quad\quad\quad\quad\quad\quad
	\text{1 + 1 dimensions}  \\[4pt]
	\sigma = 12 \, \Delta S / A
	\quad\quad\quad\quad\quad\quad
	\text{2 + 1 dimensions}  \\[4pt]
	\sigma = 48\pi \, \Delta S / A
	\quad\quad\quad\quad\quad\quad
	\text{3 + 1 dimensions}
\end{align}
The term ``area'' is used to refer to the measure of the surface of spatial co-dimension 1 in all cases, and as with other dimensionful quantities it is measured in units of the appropriate inverse power of the renormalized mass scale $m_R$.  In terms of these quantities, the dimensionless gap equation and the normalized entanglement entropy per unit area take the same simple form in any dimension:
\begin{align}
	\label{eqn:simplifiedgap}
	\varphi^2 &= f(m^2) + \frac{1}{\alpha} (m^2 - 1)  \\
	\label{eqn:simplifiedEE}
	\sigma &= - f(m^2)
\end{align}
We note the ranges of the relevant quantities in various dimensions are 
\begin{align}
	\alpha \geq 0 \quad\quad\quad  m^2 \geq 1 \quad\quad\quad f(m^2) \geq 0
	\quad\quad\quad\quad \text{1+ 1 dimensions}  \\[4pt]
	\alpha \geq 0 \quad\quad\quad m^2 \geq 1 \quad\quad\quad f(m^2) \geq 0
	\quad\quad\quad\quad \text{2+ 1 dimensions}  \\[4pt]
	\alpha \leq 0 \quad \ \ 0 \leq m^2 \leq 1 \quad \ -1 \leq f(m^2) \leq 0
	\quad\quad \text{3+ 1 dimensions}
\end{align}
The system of equations given by \eqref{eqn:simplifiedgap} and \eqref{eqn:simplifiedEE} allows us to determine the variational approximation to the entanglement entropy for fixed values of the coupling and vacuum expectation value $\varphi$.  In particular, for a fixed value of the coupling and a fixed value of $\varphi$, we solve the gap equation in Eqn. \eqref{eqn:simplifiedgap} for the variational mass $m$, and then plug $m$ into equation \eqref{eqn:simplifiedEE} to obtain the normalized entanglement entropy per unit area $\sigma$.

Before delving into a full discussion of the dependence of entanglement entropy on the coupling, we briefly point out a connection between our variational calculation of entanglement entropy in $\phi^4$ theory and a perturbative calculation.  For $d=3$, $\phi^4$ theory is widely believed to be a sick theory for finite positive coupling in a non-perturbative sense.  However, one can make sense of the theory for infinitesimal positive coupling, and treat the theory perturbatively.  To compute the entanglement entropy in the vacuum of such a theory, we consider equation \eqref{eqn:simplifiedgap} for $d=3$ and $\varphi = 0$ in the $\alpha \to 0$ limit, and find that $m^2 = 1$.  Temporarily restoring the renormalized mass scale $m_R$, we find that the solution to the gap equation in the $\alpha \to 0$ limit is $m = m_R$.  Then the approximate normalized entanglement entropy per unit area is $\sigma = - f(m_R^2) = 0$, which agrees exactly with a perturbative calculation of Hertzberg to first order in perturbation theory \cite{Hertzberg1}.  This result demonstrates that our variational framework can reproduce perturbative results in the small coupling limit.  However, we note that in the case of $\phi^4$ theory with infinitesimal positive coupling, we learn nothing about the explicit dependence of the entanglement entropy on the coupling since $\sigma = - f(m_R^2) = 0$ does not depend on the normalized coupling $\alpha$. 

\subsection{Dependence of Entanglement Entropy on the Coupling}
\label{ssec:dependence_of_EE_on_the_coupling}

Now we will analyze the dependence of the normalized entanglement entropy per unit area $\sigma$ on the coupling.  The results presented in this section demonstrate the utility of the variational techniques, and explore the relationship between entanglement and coupling.

In order to determine $\sigma$ as a function of coupling, there are two approaches.  The first approach is to solve the gap equation \eqref{eqn:simplifiedgap} at fixed coupling and fixed $\varphi$ for the variational mass $m$, and then plug $m$ into equation \eqref{eqn:simplifiedEE} to obtain $\sigma$.  In fact, the gap equation \eqref{eqn:simplifiedgap} can be solved exactly in terms of Lambert $W$ functions for $d=1$ and $d=3$, and by the quadratic formula for $d=2$.  However, the form of the corresponding normalized entanglement entropy per unit area $\sigma$ is not particularly enlightening since it is difficult to parse, although we can graph it numerically and examine the plot.  The normalized entanglement entropy per unit area $\sigma$ is plotted in each dimension on the following pages.

While we can guess some trends from the plots, it is best to be more quantitative.  A second approach to characterizing the dependence of $\sigma$ on the coupling is to compute $\left( \frac{\partial \sigma}{\partial \alpha} \right)_{\varphi}$ and $\left( \frac{\partial^2 \sigma}{\partial \alpha^2} \right)_{\varphi}$ implicitly using the gap equation, and then examine the behavior of these derivatives.  Following this approach, let us first examine $\left( \frac{\partial \sigma}{\partial \alpha} \right)_{\varphi}$.  First we note that
\begin{equation}
	\left( \frac{\partial \sigma}{\partial \alpha} \right)_{\varphi} 
	= - f'(m^2) \left( \frac{\partial m^2}{\partial \alpha} \right)_{\varphi}
\end{equation}
While we can easily compute $f'(m^2) = \partial f / \partial m^2$ using the explicit form of $f$ in each dimension, we need to use the gap equation to compute $\left( \frac{\partial m^2}{\partial \alpha} \right)_{\varphi}$.  Considering $m$ as a function of the scalar field and coupling given by the gap equation, and differentiating the dimensionless gap equation with $\varphi$ held fixed leads to the general result
\begin{equation}
\label{m^2_first_derivative}
	\left( \frac{\partial m^2}{\partial \alpha} \right)_{\varphi} = \frac{m^2 - 1}{\alpha^2} \left( \frac{1}{\alpha} + f'(m^2) \right)^{-1}
\end{equation}
Specifically, in each dimension,
\begin{align}
	\frac{\partial m^2}{\partial \alpha} &= \frac{m^2 - 1}{\alpha \ (1 + \alpha / m^2)}
	\quad\quad\quad\quad\quad\quad\quad\quad 
	\text{1 + 1 dimensions}  \\[4pt]
	\frac{\partial m^2}{\partial \alpha} &= \frac{m^2 - 1}{\alpha \, (1 + \alpha / 2 m)}
	\quad\quad\quad\quad\quad\quad\quad\quad
	\text{2 + 1 dimensions}  \\[4pt]
\label{m^2derivative3+1}
	\frac{\partial m^2}{\partial \alpha} &= \frac{m^2 - 1}{\alpha \, (1+ \alpha \log m^2)}
	\quad\quad\quad\quad\quad\quad\quad
	\text{3 + 1 dimensions}
\end{align}
Calculation of the first derivative of the normalized entanglement entropy per unit area $\sigma$ is now simple:
\begin{equation}
\label{sigma_first_derivative}
	\left( \frac{\partial \sigma}{\partial \alpha} \right)_{\varphi} 
	= - f'(m^2) \left( \frac{\partial m^2}{\partial \alpha} \right)_{\varphi}
	= - \frac{(m^2 - 1)}{\alpha^2} \left( 1 + \frac{1}{\alpha f'(m^2)} \right)^{-1}
\end{equation}
Using the ranges of the relevant quantities noted above, we find
\begin{align}
\label{monotonicity1+1}
	\left( \frac{\partial \sigma}{\partial \alpha} \right)_{\varphi}  &\leq 0
	\quad\quad\quad\quad\quad
	\text{1 + 1 dimensions}  \\[4pt]
\label{monotonicity2+1}
	\left( \frac{\partial \sigma}{\partial \alpha} \right)_{\varphi} &\leq 0
	\quad\quad\quad\quad\quad
	\text{2 + 1 dimensions}  \\[4pt]
\label{monotonicity3+1,1}
	\left( \frac{\partial \sigma}{\partial \alpha} \right)_{\varphi} &\leq 0
	\quad\quad\quad 	
	\text{3 + 1 dimensions, $m^2 \geq e^{1/\alpha}$ for $\alpha \leq 0$}  \\[4pt]
%\label{monotonicity3+1,1.5}
%	\left( \frac{\partial \sigma}{\partial \alpha} \right)_{\varphi} &\geq 0
%	\quad\quad\quad 	
%	\text{3+ 1 dimensions, $m^2 \geq e^{1/\alpha}$ for $\alpha \geq 0$}  \\[4pt]
\label{monotonicity3+1,2}
	\left( \frac{\partial \sigma}{\partial \alpha} \right)_{\varphi} &\geq 0
	\quad\quad\quad 	
	\text{3 + 1 dimensions, $m^2 \leq e^{1/\alpha}$ for $\alpha \leq 0$}
%\label{monotonicity3+1,2.5}
%	\left( \frac{\partial \sigma}{\partial \alpha} \right)_{\varphi} &\leq 0
%	\quad\quad\quad 	
%	\text{3+ 1 dimensions, $m^2 \leq e^{1/\alpha}$ for $\alpha \geq 0$}
\end{align}
We also note the results for \emph{positive} coupling $\alpha$ in 3+1 dimensions, which are related to the negative coupling results by the duality relation we discussed previously
\begin{align}
\label{monotonicity3+1,1.5}
	\left( \frac{\partial \sigma}{\partial \alpha} \right)_{\varphi} &\geq 0
	\quad\quad\quad 	
	\text{3 + 1 dimensions, $m^2 \geq e^{1/\alpha}$ for $\alpha \geq 0$}  \\[4pt]
\label{monotonicity3+1,2.5}
	\left( \frac{\partial \sigma}{\partial \alpha} \right)_{\varphi} &\leq 0
	\quad\quad\quad 	
	\text{3 + 1 dimensions, $m^2 \leq e^{1/\alpha}$ for $\alpha \geq 0$}
\end{align}

We will now examine the results in equations (\ref{monotonicity1+1}) through (\ref{monotonicity3+1,2}).  First, for equations (\ref{monotonicity1+1}) and (\ref{monotonicity2+1}), we note that
\begin{equation}
\label{DeltaS_to_S_1}
\left( \frac{\partial \sigma}{\partial \alpha} \right)_{\varphi} = \frac{12}{A} \left( \frac{\partial \Delta S}{\partial \alpha} \right)_{\varphi} = \frac{12}{A}\left( \frac{\partial S}{\partial \alpha} \right)_{\varphi} \quad\quad\quad
	\text{1+1 and 2+1 dimensions}
\end{equation}
In the above equation, we can replace $\Delta S$ with $S$ since the divergent pieces of $S$ have no dependence on the coupling for $d=1$ and $d=2$.  Thus, while $S$ is divergent and regularization dependent, $\left( \frac{\partial S}{\partial \alpha} \right)_{\varphi}$ is finite and independent of regularization, and so represents a meaningful physical quantity.  On the other hand, for precarious $\phi^4$ theory where $d=3$, $\left( \frac{\partial S}{\partial \alpha} \right)_{\varphi}$ is not finite and so is not proportional to $\left( \frac{\partial \sigma}{\partial \alpha} \right)_{\varphi}$.  However, the divergent term in  $\left( \frac{\partial S}{\partial \alpha} \right)_{\varphi}$ is proportional to $\frac{\partial m^2}{\partial \alpha} \log(\epsilon)$ and so in light of equation (\ref{m^2derivative3+1}), for small $\epsilon$ we find that $\left( \frac{\partial S}{\partial \alpha} \right)_{\varphi}$ has the same sign as $\left( \frac{\partial \sigma}{\partial \alpha} \right)_{\varphi}$.  So to analyze the monotonicity of $S$ for precarious $\phi^4$ theory, it suffices to analyze the monotonicity of $\sigma$. 

In equations (\ref{monotonicity1+1}) and (\ref{monotonicity2+1}), we see that for $d=1$ and $d=2$ the variational approximation to the entanglement entropy is monotonically \textit{decreasing} with respect to coupling for fixed $\varphi$.  This result is surprising, since one might have expected the entanglement entropy to increase as the coupling is increased.  We can trust our result in a neighborhood of zero coupling for which the variational approximation is accurate.  Thus, a conservative interpretation of equations (\ref{monotonicity1+1}) and (\ref{monotonicity2+1}) is that for $d=1$ and $d=2$ the entanglement entropy is monotonically decreasing in a neighborhood of zero coupling.  More ambitiously, we might take the monotonicity of the variational approximations of the entanglement entropies as an indication that the \textit{true} entanglement entropies for $d=1$ and $d=2$ are monotonically decreasing for all values of the coupling.  This amounts to little more than a conjecture at the moment, but is nonetheless suggestive.   

Now we turn to the precarious $\phi^4$ theory for $d=3$, and consider equations (\ref{monotonicity3+1,1}), (\ref{monotonicity3+1,1.5}), (\ref{monotonicity3+1,2}) and (\ref{monotonicity3+1,2.5}).  For $m^2 \geq e^{1/\alpha}$, the theory is stable and asymptotically free and equation (\ref{monotonicity3+1,1}) indicates that for $\alpha \leq 0$, $\sigma$ is monotonically \textit{increasing} for fixed $\varphi$ as we increase the magnitude of the negative coupling (i.e., make the coupling more negative).  Similarly for $m^2 \geq e^{1/\alpha}$, equation (\ref{monotonicity3+1,1.5}) indicates that for $\alpha \geq 0$, $\sigma$ is monotonically \textit{increasing} for fixed $\varphi$ as we increase the magnitude of the positive coupling.  For $m^2 \leq e^{1/\alpha}$ the theory is unstable, and interestingly equation (\ref{monotonicity3+1,2}) indicates that for such an unstable theory when $\alpha \leq 0$, $\sigma$ is monotonically \textit{decreasing} for fixed $\varphi$ as we increase the magnitude of the negative coupling.  Likewise for the unstable theory, equation (\ref{monotonicity3+1,2.5}) tell us that for $\alpha \geq 0$ we have that $\sigma$ is monotonically \textit{increasing} for fixed $\varphi$ as we increase the magnitude of the positive coupling.  Thus, it appears that the direction of the monotonicity of $\sigma$ is tied to the stability of the theory.  While we understand that the variational approximation is accurate for small values of $|\alpha|$, we might trust in the variational approximation and conjecture that the sign of the derivative of an \textit{exact} entanglement entropy with respect to coupling depends on the stability of the QFT in question.

\newpage
Having examined the first derivative $\left( \frac{\partial \sigma}{\partial \alpha} \right)_{\varphi}$ to analyze the monotonicity of the entanglement entropy with respect to coupling, we now turn to $\left( \frac{\partial^2 \sigma}{\partial \alpha^2} \right)_{\varphi}$ to analyze the convexity of the entanglement entropy with respect to coupling for a fixed value of the expectation value of the scalar field.  Using equations (\ref{m^2_first_derivative}) and (\ref{sigma_first_derivative}), we find that the second derivative of the normalized entanglement entropy per unit area $\sigma$ is
\bigskip
\begin{equation}
	\left( \frac{\partial^2 \sigma}{\partial \alpha^2} \right)_{\varphi} 
	= \frac{2(m^2-1)(f'(m^2))^2}{\alpha\left(1 + \alpha \, f'(m^2)\right)^{2}}
	- \frac{(m^2-1)^2 f''(m^2)}{\alpha^2 \left(1 + \alpha \, f'(m^2)\right)^{3}}
\end{equation}
Again using the ranges of the relevant quantities noted above, we have
\begin{align}
\label{convexity1+1}
	\left( \frac{\partial^2 \sigma}{\partial \alpha^2} \right)_{\varphi}  &\geq 0
	\quad\quad\quad\quad\quad \,
	\text{1 + 1 dimensions}  \\[4pt]
\label{convexity2+1}
	\left( \frac{\partial^2 \sigma}{\partial \alpha^2} \right)_{\varphi} &\geq 0
	\quad\quad\quad\quad\quad
	\text{2 + 1 dimensions}  \\[4pt]
\label{convexity3+1,1}
	\left( \frac{\partial^2 \sigma}{\partial \alpha^2} \right)_{\varphi} &\geq 0
	\quad\quad\quad 	
	\text{3 + 1 dimensions, $m^2 \geq e^{1/\alpha}$ for all $\alpha$}  \\[4pt]
\label{convexity3+1,2}
	\left( \frac{\partial^2 \sigma}{\partial \alpha^2} \right)_{\varphi} &\leq 0
	\quad\quad\quad 	
	\text{3 + 1 dimensions, $m^2 \leq e^{1/\alpha}$ for all $\alpha$}
\end{align}
Similar to equation (\ref{DeltaS_to_S_1}), we note that
\bigskip
\begin{equation}
\label{DeltaS_to_S_2}
	\left( \frac{\partial^2 \sigma}{\partial \alpha^2} \right)_{\varphi} 
	= \frac{12}{A} \left( \frac{\partial^2 \Delta S}{\partial \alpha^2} \right)_{\varphi} 
	= \frac{12}{A}\left( \frac{\partial^2 S}{\partial \alpha^2} \right)_{\varphi}
	\quad\quad\quad \text{1 + 1 and 2 + 1 dimensions}
\end{equation}
and so $\left( \frac{\partial^2 S}{\partial \alpha^2} \right)_{\varphi}$ is a finite and regularization independent quantity for $d=1$ and $d=2$.  By equations (\ref{convexity1+1}) and (\ref{convexity2+1}), for $d=1$ and $d=2$ we have that $\sigma$ is convex in the coupling for fixed $\varphi$.  Since for $d=1$ and $d=2$, $\sigma$ is monotonically decreasing, as the coupling increases the entanglement decreases less and less rapidly.

Now let us examine precarious $\phi^4$ theory where $d=3$.  For this theory, $\left( \frac{\partial^2 S}{\partial \alpha^2} \right)_{\varphi}$ is not finite and is not proportional to $\left( \frac{\partial^2 \sigma}{\partial \alpha^2} \right)_{\varphi}$.  In fact, the divergent term in $\left( \frac{\partial^2 S}{\partial \alpha^2} \right)_{\varphi}$ is proportional to $\frac{\partial^2 (m^2)}{\partial \alpha^2} \log(\epsilon)$.  Since for $d=3$ we have
\bigskip
\begin{equation}
	\frac{\partial^2 (m^2)}{\partial \alpha^2} 
	= - \frac{(m^2 - 1)}{\alpha \, m^2 (1 + \alpha \log(m^2))^3} \left(m^2 \left(2 \alpha \log(m^2)^2 + 2\log(m^2) + 1 \right) - 1 \right)
\end{equation}
for small $\epsilon$ we see that $\left( \frac{\partial^2 S}{\partial \alpha^2} \right)_{\varphi}$ has the same sign as $\left( \frac{\partial^2 \sigma}{\partial \alpha^2} \right)_{\varphi}$.  Therefore, to analyze the convexity of $S$ for precarious $\phi^4$ theory it is sufficient to analyze the convexity of $\sigma$. 

By equation (\ref{convexity3+1,1}), in precarious $\phi^4$ theory for the stable region $m^2 \geq e^{1/\alpha}$ we have that $\sigma$ is convex in the coupling for fixed $\varphi$.  However, by equation (\ref{convexity3+1,2}), in the unstable region where $m^2 \leq e^{1/\alpha}$ we have that $\sigma$ is concave in the coupling for fixed $\varphi$.  We conclude that in the variational approximation, the convexity of the entanglement entropy is linked with the stability of the theory.  One might conjecture that this connection between convexity and stability holds for exact calculations of the entanglement entropy.

\bigskip

\begin{figure}[hp]
	\begin{centering}
	\scalebox{0.9}{\includegraphics[width=1.2\textwidth,center]{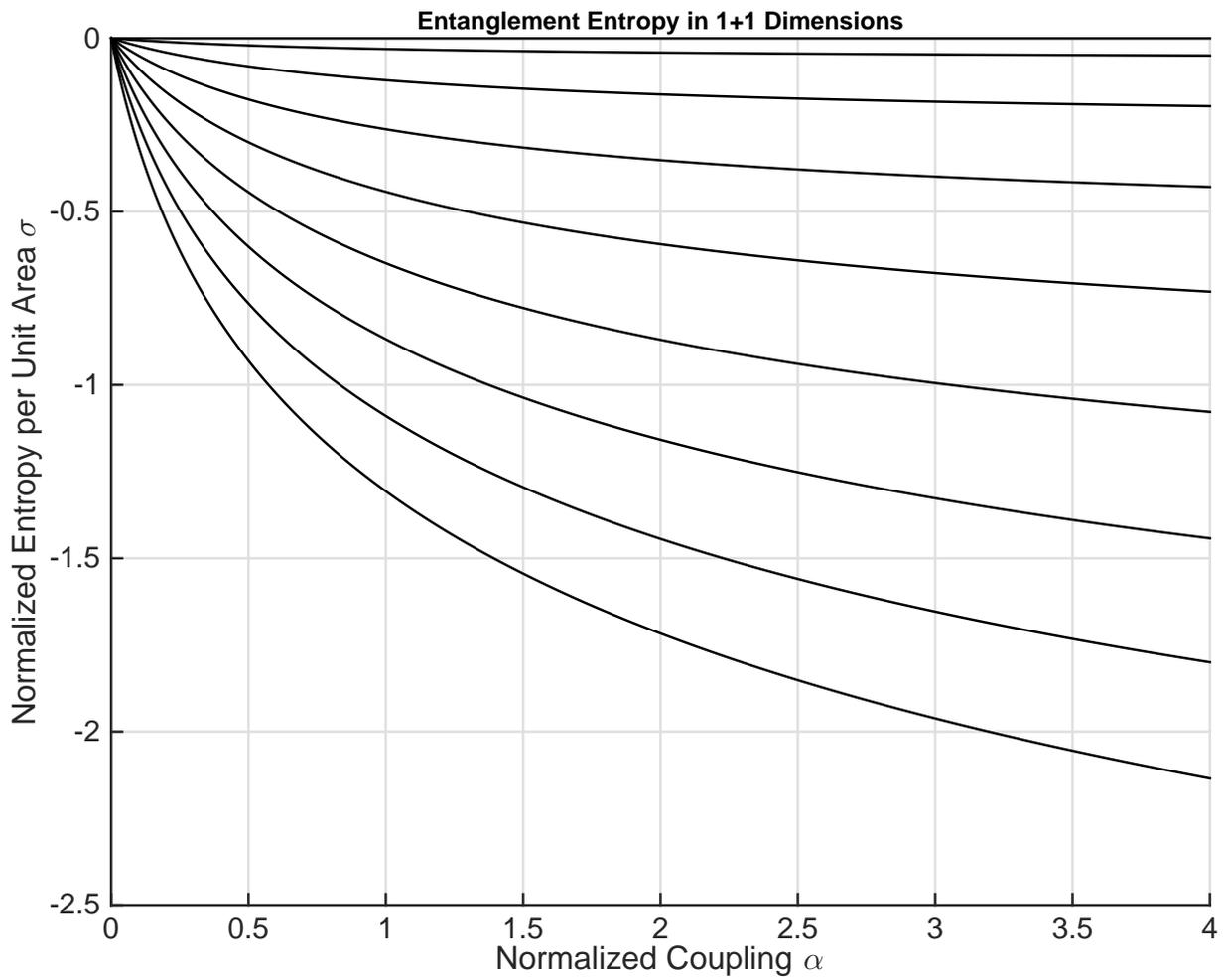}}
	\end{centering}
	\caption{Normalized entanglement entropy per unit area in 1+1 dimensions as a function of the normalized coupling $\alpha$, for various values of the normalized scalar field $\varphi$.  Lower lines correspond to higher values of the scalar field.}
	\label{fig:ee1d}
\end{figure}

\begin{figure}[hp]
	\begin{centering}
	\scalebox{0.9}{\includegraphics[width=1.2\textwidth,center]{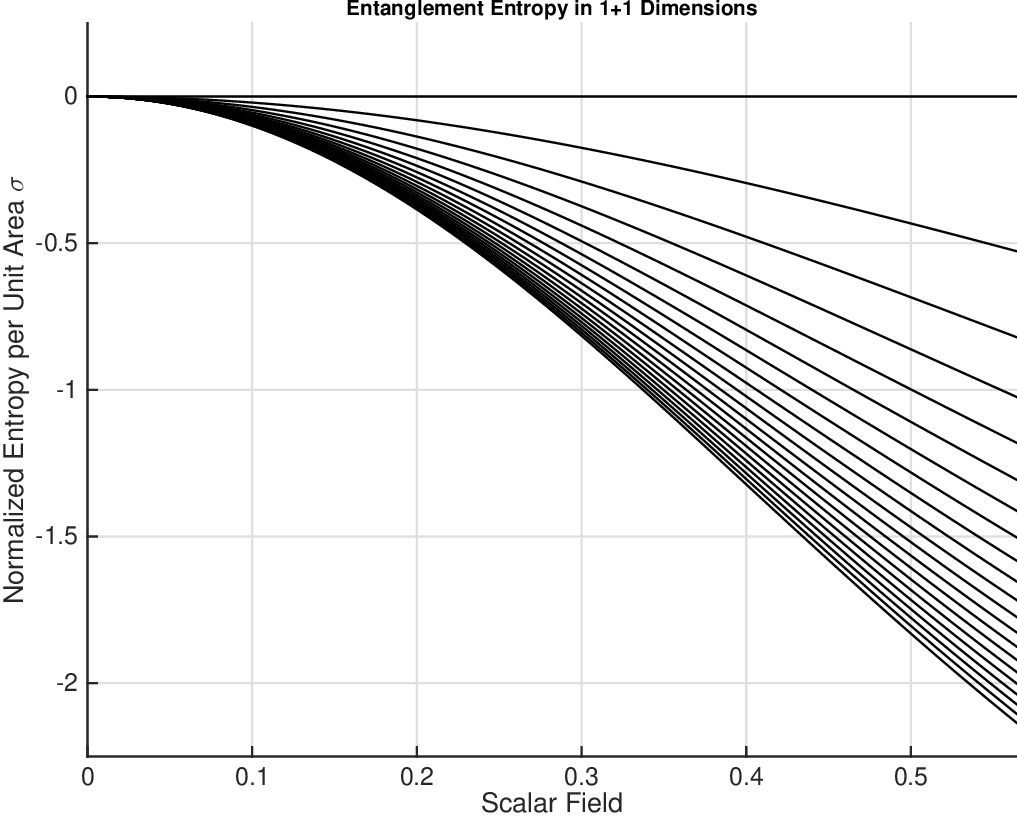}}
	\end{centering}
	\caption{Normalized entanglement entropy per unit area in 1+1 dimensions as a function of the normalized scalar field $\varphi$, for various values of the normalized coupling $\alpha$.  Lower lines correspond to stronger couplings.}
	\label{fig:ee1d_2}
\end{figure}

\begin{figure}[hp]
	\begin{centering}
	\scalebox{0.9}{\includegraphics[width=1.2\textwidth,center]{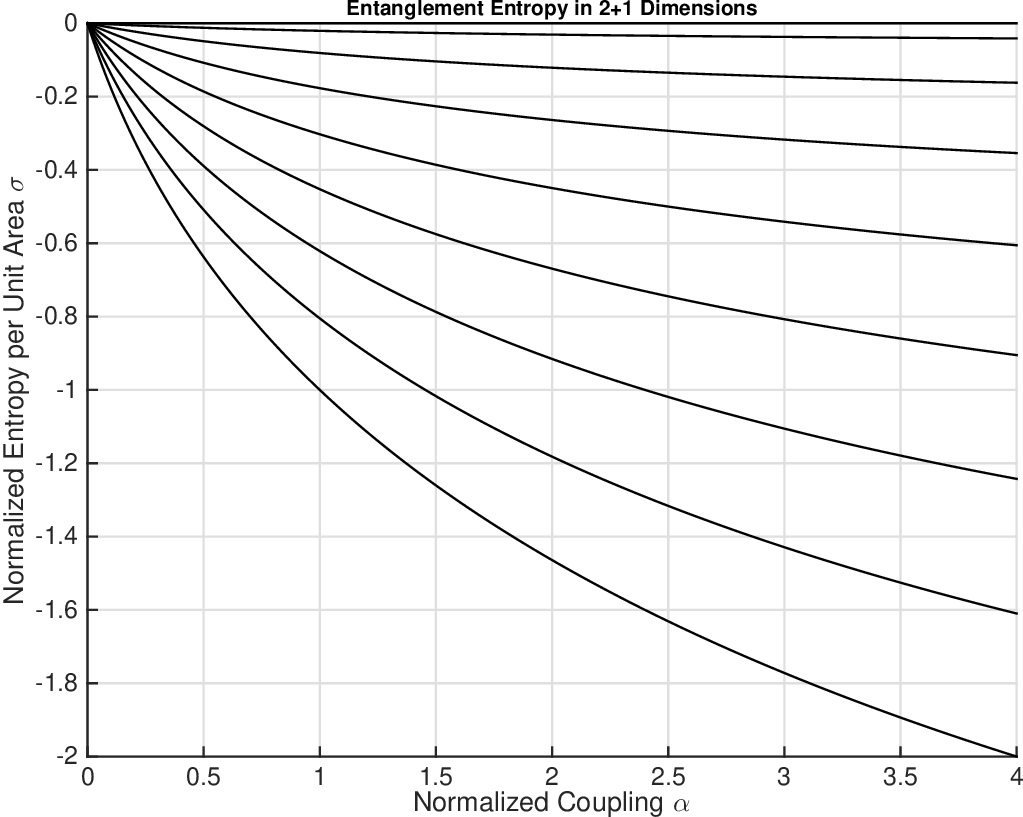}}
	\end{centering}
	\caption{Normalized entanglement entropy per unit area in 2+1 dimensions as a function of the normalized coupling $\alpha$, for various values of the normalized scalar field $\varphi$.    Lower lines 	correspond to higher values of the scalar field.}
	\label{fig:ee2d}
\end{figure}

\begin{figure}[hp]
	\begin{centering}
	\scalebox{0.9}{\includegraphics[width=1.2\textwidth,center]{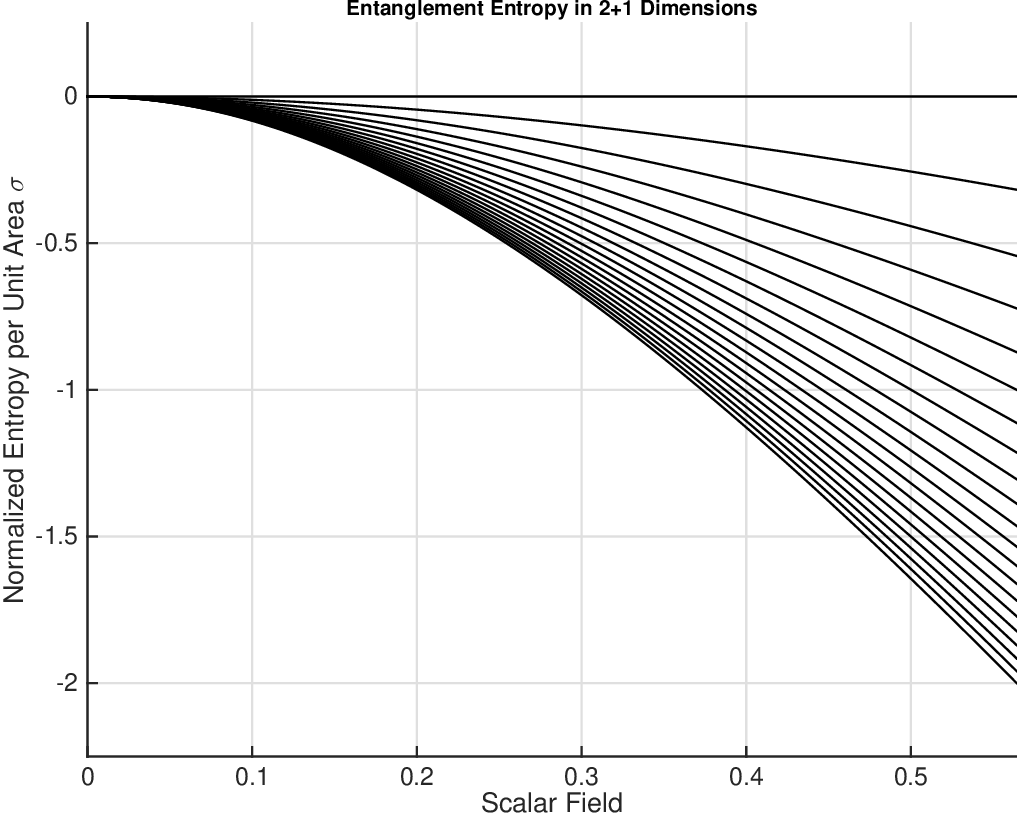}}
	\end{centering}
	\caption{Normalized entanglement entropy per unit area in 2+1 dimensions as a function of the normalized scalar field $\varphi$, for various values of the normalized coupling $\alpha$.  Lower lines correspond to stronger couplings.}
	\label{fig:ee2d}
\end{figure}

\begin{figure}[hp]
	\begin{centering}
	\scalebox{0.9}{\includegraphics[width=1.2\textwidth,center]{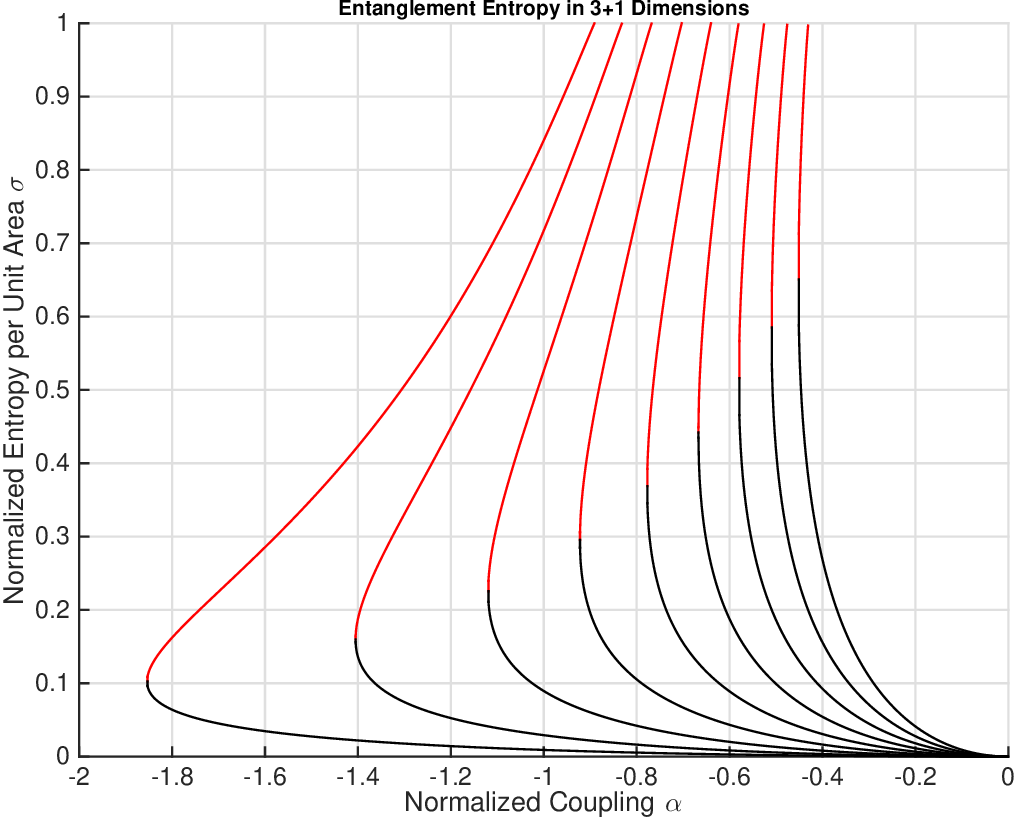}}
	\end{centering}
	\caption{Normalized entanglement entropy per unit area in 3+1 dimensions, as a function of the normalized coupling $\alpha < 0$, for various values of the normalized scalar field $\varphi$.  The upper branches (in red) correspond to unstable solutions of the gap equation which do not minimize the Gaussian Effective Potential; the lower branches (in black) correspond to stable or metastable solutions of the gap equation.  Curves from left to right correspond to increasing values of the scalar field.}
	\label{fig:ee3d}
\end{figure}

\begin{figure}[hp]
	\begin{centering}
	\scalebox{0.9}{\includegraphics[width=1.2\textwidth,center]{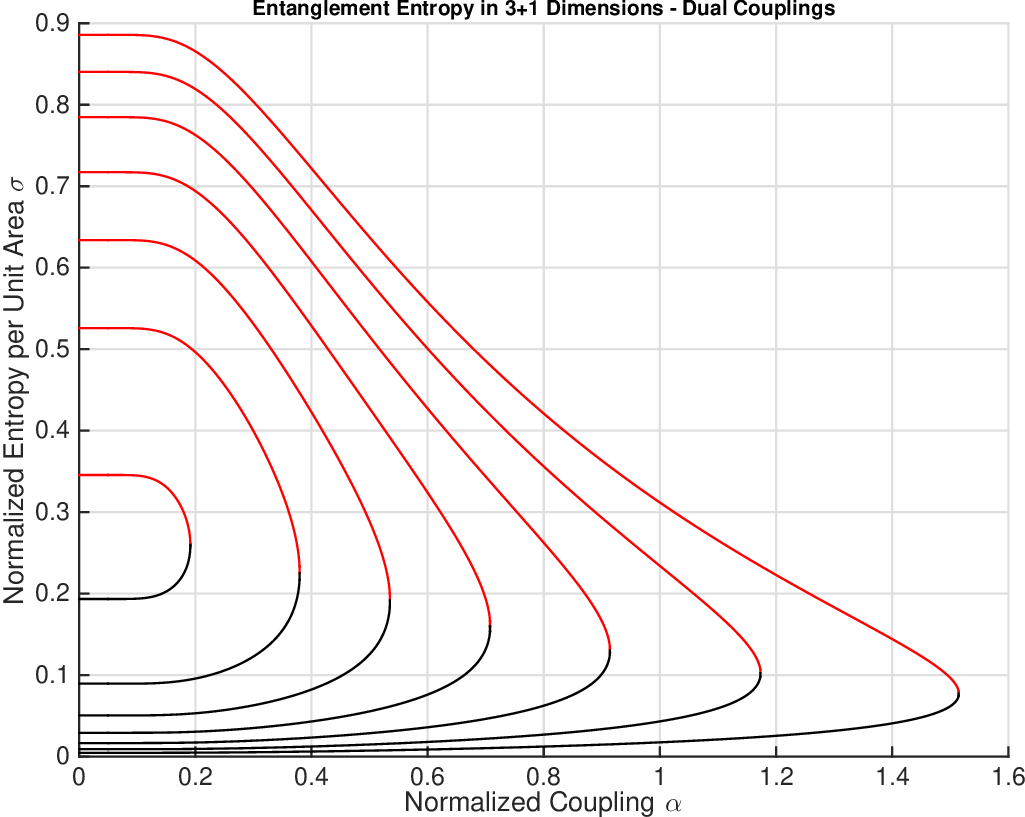}}
	\end{centering}
	\caption{Normalized entanglement entropy per unit area in 3+1 dimensions, as a function of the normalized dual coupling $\alpha > 0$, for various values of the normalized scalar field $\varphi$.  Again, the upper branches (in red) correspond to unstable solutions of the gap equation which do not minimize the Gaussian Effective Potential; the lower branches (in black) correspond to stable or metastable solutions of the gap equation.  Curves from left to right correspond to decreasing values of the scalar field.}
	\label{fig:ee3d_dual_coupling}
\end{figure}

\begin{figure}[hp]
	\begin{centering}
	\scalebox{0.9}{\includegraphics[width=1.2\textwidth,center]{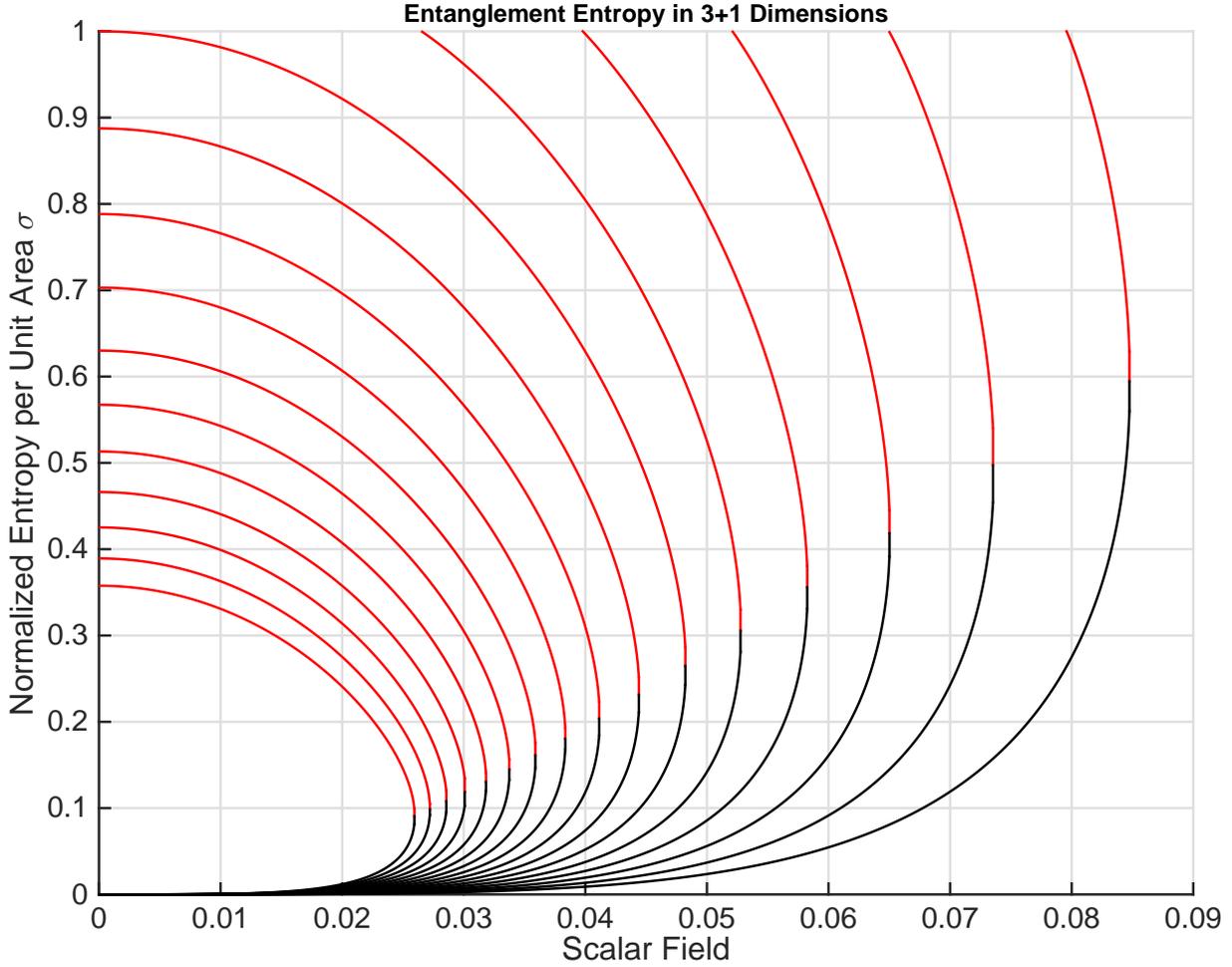}}
	\end{centering}
	\caption{Normalized entanglement entropy per unit area in 3+1 dimensions for precarious $\phi^4$ theory, as a function of the normalized scalar field $\varphi$, for various values of the normalized coupling $\alpha$.  Couplings in this plot take only negative values; the coupling strength decreases in magnitude (becomes less negative and approaches zero) from left to right, or from inner to outer curves.  The upper branches (in red) correspond to unstable solutions of the gap equation which do not minimize the Gaussian Effective Potential; the lower branches (in black) correspond to stable or metastable solutions of the gap equation.}
	\label{fig:ee3d_2}
\end{figure}

\newpage
\section{Renormalization Group Flow of Entanglement Entropy}
\label{sec:RGflowofEE}
In this section we compute the renormalization group equation for the normalized entanglement entropy per unit area $\sigma = - f(m^2/\mu^2)$, and show that it is simply related to the usual beta function for the renormalized coupling. For 1+1 and 2+1 dimensional $\phi^4$ theory, as discussed earlier, there is merely a finite reparameterization of the coupling, with no coupling renormalization depending on a renormalization mass scale $\mu$, so the beta function vanishes.  However, for precarious $\phi^4$ theory in 3+1 dimensions, there is a non-vanishing beta function.  If we differentiate the coupling renormalization equation \eqref{eqn:coupling_renormalization} with respect to the renormalization scale $\mu$ for fixed bare coupling and fixed cutoff, then in terms of the normalized coupling $\alpha = \lambda_R / 96 \pi^2$ the beta function is given by
\begin{equation}
\label{betafunction1}
	\beta(\alpha) = \mu \frac{d\alpha}{d\mu} = 2 \alpha^2
\end{equation}
Since for precarious $\phi^4$ theory we have $\alpha < 0$, the beta function in equation (\ref{betafunction1}) tells us that the theory is asymptotically free.  In particular, $1/\alpha \sim \log \mu^2$.

Using the simplified forms of the gap equation and entanglement entropy formulas in equations \eqref{eqn:simplifiedgap} and \eqref{eqn:simplifiedEE}, we obtain the renormalization group equation for the normalized entanglement entropy per unit area
\begin{equation}
\label{EERG}
\mu \frac{d\sigma}{d\mu} = \frac{\left(1 - \frac{m^2}{\mu^2} \right) f'(m^2/\mu^2)}{\left[\frac{1}{\alpha} + f'(m^2/\mu^2) \right]} \, \frac{\mu}{\alpha^2} \frac{d\alpha}{d\mu}
= \frac{\left(1 - \frac{m^2}{\mu^2} \right) f'(m^2/\mu^2)}{\left[\frac{1}{\alpha} + f'(m^2/\mu^2) \right]} \, \frac{1}{\alpha^2} \, \beta(\alpha)
\end{equation}
We will refer to this quantity as the \emph{entanglement entropy beta function}, although any such quantity obviously depends on the geometry of the region that defines the entanglement entropy as well as the quantum field theory itself.  In 1+1 and 2+1 dimensional $\phi^4$ theory, since the beta function vanishes, the renormalization group equation for the normalized entanglement entropy per unit area is trivial.  Simplifying equation (\ref{EERG}) in the case of precarious $\phi^4$ theory, we obtain
\begin{equation}
\label{EERG3}
\mu \frac{d\sigma}{d\mu} = - 2 \, \frac{(1 - \frac{m^2}{\mu^2}) \log \frac{m^2}{\mu^2}}{ \frac{1}{\alpha} - \log \frac{m^2}{\mu^2}}
\end{equation}
Since $m^2/\mu^2 \leq 1$, this entanglement entropy beta function for precarious $\phi^4$ theory is ``asymptotically free'', meaning that $\mu \frac{d\sigma}{d\mu} \le 0$, but only in the stable region of precarious $\phi^4$ where $m^2/\mu^2 > e^{1/\alpha}$.  Equality $\mu \frac{d\sigma}{d\mu} = 0$ holds only in the vacuum where $m^2/\mu^2 = 1$.  The sign is reversed for the unstable region, with singular behavior when $m^2/\mu^2 = e^{1/\alpha}$ where precarious $\phi^4$ theory goes from being stable to unstable.  It will be interesting to see which properties of the entanglement entropy beta function for precarious $\phi^4$ are shared by other theories with asymptotic freedom, such as the Gross-Neveu model and non-abelian gauge theories.

\section{Conclusions}
\label{sec:conclusions}

We have applied variational methods to approximate entanglement entropies for scalar $\phi^4$ theories.  Within our approximation scheme we have found that in $d=1$ and $d=2$ dimensions, the entanglement entropy is monotonically decreasing and convex with respect to the coupling.  This result is unexpected since the authors anticipated that larger coupling would lead to more entanglement.  Our results are accurate for small coupling, although the approximation scheme is well-defined for all values of the coupling.  Inspired by the variational results, we conjecture that the \textit{exact} entanglement entropy in $d=1$ and $d=2$ is monotonically decreasing and convex with respect the coupling for the entire range of positive couplings.  Indeed, it would be strange if the derivative of the entanglement entropy switched signs for some particular value of the coupling.  If this were the case, then this value of the coupling would be of immense physical interest since it would correspond to a local minimum of entanglement entropy.

For precarious $\phi^4$ theory in the stable range, we have also demonstrated that the variational approximation to the entanglement entropy is monotonic in $\alpha$ for $\alpha \leq 0$ and for the dual values $\alpha \geq 0$, and is convex for all $\alpha$.  This theory is of particular interest since it is asymptotically free.  As we tune the parameters of the theory so that it becomes unstable, the derivative of the variational approximation to the entanglement entropy changes sign exactly at the transition between quasi-stability and instability.  Likewise, the second derivative changes sign.  We have also shown the the ``entanglement entropy beta function" for precarious $\phi^4$ theory in 3+1 dimensions is asymptotically free.

%We have also shown that the positive coupling region of $\phi^4$ theory which is well-defined non-perturbatively in our variational scheme also has an approximate entanglement entropy which is monotonic and convex with respect to the coupling.  This result was obtained via a duality with precarious $\phi^4$ theory.

It would be interesting to see whether or not monotonicity and convexity with respect to coupling are generic features of entanglement entropy for QFT's, and to explore why certain theories appear to have entanglement entropies which are monotonically \textit{decreasing} with respect to coupling.  It would also be interesting to examine the entanglement entropy beta functions of asymptotically free theories, and see if the behavior of precarious $\phi^4$ theory is generic.  In future work, we plan to apply our techniques to quantum field theories with fermions as well as bosons.  In addition, we hope to explore the entanglement structure of gauge theories with variational techniques.  This would be a valuable step towards analyzing entanglement entropies of quantum field theories appearing in the Standard Model.

\newpage
\section*{Acknowledgments}
\label{sec:ack}
The authors would like to thank Arthur Kerman, Mark Hertzberg, Frank Wilczek, Hong Liu, Julian Sonner, Roman Jackiw, Jeffrey Goldstone, Brian Swingle, John McGreevy, Eric Tonni, and Paul Stevenson for valuable discussions and feedback.

\noindent
Jordan Cotler is supported by the Fannie and John Hertz Foundation and the Stanford Graduate Fellowship program.  Mark Mueller is supported by the U.S. Department of Energy under grant Contract Number DE-SC00012567.

\end{document}